\documentclass[
superscriptaddress,
nofootinbib,
 amsmath,amssymb,
 aps,
pra,
twocolumn,
10pt,
]{revtex4-2}

\usepackage{natbib}
\usepackage{graphicx} 
\usepackage{dcolumn}
\usepackage{bm}
\usepackage{xcolor}
\usepackage{qcircuit}
\usepackage{braket}
\graphicspath{ {figures/} }
\usepackage[colorlinks=true,citecolor=blue]{hyperref}
\usepackage{amsmath}
\usepackage{amssymb}
\usepackage{bbm}
\usepackage[thinc]{esdiff}
\usepackage{svg}
\usepackage[normalem]{ulem}
\usepackage{enumitem}
\usepackage{comment}

\renewcommand{\figureautorefname}{Fig.~\negthinspace}
\renewcommand{\sectionautorefname}{Sec.~\negthinspace}
\renewcommand{\equationautorefname}{Eq.~\negthinspace}
\renewcommand{\appendixautorefname}{App.~\negthinspace}

\newcommand{\Tr}{\mathrm{Tr}}

\begin{document}

\title{Performance limits of a quantum receiver for detecting phase-modulated communication signals}

\author{William M. Watkins}
\email{wwatki11@jhu.edu}
\affiliation{%
 Johns Hopkins Applied Physics Laboratory, Laurel, Maryland 20723
}%
\affiliation{
    Department of Physics \& Astronomy, Johns Hopkins University, Maryland 21218
}

\author{Leigh Norris}
\altaffiliation[]{Present address: Quantinuum, Broomfield, Colorado, USA}
\affiliation{%
 Johns Hopkins Applied Physics Laboratory, Laurel, Maryland 20723
}

\author{Paraj Titum}
\affiliation{%
 Johns Hopkins Applied Physics Laboratory, Laurel, Maryland 20723
}%
\affiliation{
    Department of Physics \& Astronomy, Johns Hopkins University, Maryland 21218
}

\date{\today}

\begin{abstract}

Quantum sensors are an ideal candidate for detecting weak electromagnetic signals because of their exceptional sensitivity and compact form factor.
In this work, we analyze the performance of a quantum-sensor-based receive chain for demodulating information encoded in phase-modulated electromagnetic waves. We introduce a generalized cumulant expansion to model a noisy quantum receiver and use it to compare the performance of various quantum demodulation protocols. Employing bit error probability (BEP) and channel capacity as quantitative performance metrics, we compare the capabilities of ensembles of quantum sensors—both unentangled and entangled—using Binary Phase-Shift Keying (BPSK) as a representative example of phase modulation. 
We identify conditions when the channel capacity of an ensemble of quantum sensors may surpass the limits of a classical electrically small antenna. 
Additionally, we discuss modifications to the quantum protocol that enables high-fidelity data recovery even in the presence of sensor noise and channel distortions. Finally, we explore practical performance limits of such a quantum receive chain, with a focus on NV-diamond as the quantum sensor platform.
\end{abstract}

\maketitle

\section{Introduction}

Quantum sensing promises to utilize the strong sensitivity of quantum systems to their environment to attain unparalleled precision for various metrological applications~\cite{Degen_2017, Kantsepolsky_2023, Aslam_2023}; e.g., measuring small electric or magnetic fields \cite{Kitching_2011, Esat_2024}, timekeeping \cite{K_m_r_2014, Grotti_2018, Beloy_2021, Malia_2022}, and gravimetry \cite{Bothwell_2022, Stray_2022}. There are a wide variety of quantum sensing applications depending on the sensor modality; e.g., superconducting qubits to study magnons~\cite{Lachance_Quirion_2020, Kristen_2020, Wolski_2020, danilin2022quantum, Schultz:21}, Rydberg atoms to measure electric fields~\cite{Adams_2019,gilmore2021,Fancher_2021, schmidt2023rydberg, prajapati2022tv, Holloway_2022}, and nitrogen-vacancy~(NV) centers in diamond to measure magnetic fields~\cite{Kim_2019, Rembold_2020, doi:10.1126/sciadv.adg2080} to name a few.
From the perspective of estimation theory, quantum sensing may be viewed as a quantum parameter estimation problem---interrogating a quantum state to estimate an unknown parameter. The quantum Cram{\'e}r-Rao bound provides a lower bound on the uncertainty in the parameter estimate using an ensemble of quantum systems~\cite{Rao1992, Braunstein_1994}. According to this bound, the ultimate precision achievable for an ensemble of $N$ unentangled quantum sensors falls as $\sim N^{-1/2}$; also known as the standard quantum limit~(SQL). However, if the ensemble is entangled, the SQL can be surpassed; the minimum achievable precision falls as $\sim N^{-1}$, which is also known as the Heisenberg limit~(HL)~\cite{Giovannetti_2004, Nagata_2007}. These limits play a crucial role in quantifying the potential advantages of quantum sensing over classical methods~\cite{Zwierz_2010, Tsujino_2011, Napolitano_2011, Demkowicz_Dobrza_ski_2012, Chaves_2013, Kura_2020}.

Recently, quantum sensors have been used for various applications in classical radio-frequency (RF) communications. For example, Rydberg atom-based vapor cells have been demonstrated as a compact receiver for amplitude, frequency and phase modulated signals,~\cite{Holloway_2014, Holloway_2019, Simons_2019, berweger2023phaseresolved}.
The Rydberg-atom-based antenna utilizes Electromagnetic Induced Transparency (EIT), 
in which the applied RF signal effects a change in the vapor cell via the Autler-Townes (AT) splitting~\cite{Fleischhauer_2005, Cox_2011}.
While this sensitivity to signal amplitude can be leveraged to detect modulated RF signals, the performance of this antenna system is limited by the time it takes for the AT splitting to emerge, which ultimately limits the data rates~\cite{Sedlacek_2013, Prajapati_2022, Bohaichuk_2023}. 

Nitrogen‐Vacancy (NV) centers in diamond are a particularly promising candidate for spin-based quantum sensing, because they host an electron spin with long coherence times at room temperature~\cite{Balasubramanian2009, Bar-Gill2012, Herbschleb2019, Morishita2020}, and allow for all-optical initialization and readout of the spin state.
Microwave and RF control fields target and drive coherent transitions between spin levels, enabling preparation of superposition states and making the spins sensitive to weak, time-varying magnetic fields.
There have been numerous demonstrations using NV centers in magnetometry~\cite{zhengzhi_2023, Cox2018, Barry_2024, Wang2021, Lamba_2024, Bürgler2023}, including as RF signal analyzers~\cite{Chipaux_2015, Magaletti_2022, Zhang_J_2023}, as well as quantum frequency mixers~\cite{gerginov_2017, wang2022, Attrash_2023, Yu_2023}. 
However, practical use as a receiver is constrained by the finite coherence time from spin dephasing. To assess and optimize NV-based quantum receivers, accurate modeling of these noise processes is essential.

In this study, we investigate an alternative implementation of a quantum receiver that relies on the angular precession of an ensemble of spin-based quantum sensors rather than the EIT-based approach demonstrated in Rydberg atom ensembles.
We develop a demodulation protocol that utilizes quantum control~\cite{Titum_2021} to tailor the dynamics of these spin ensembles, and derive analytical expressions for the the bit error probability (BEP) and channel capacity, which quantifies the error in demodulation. 
Our analysis builds upon quantum state discrimination theory \cite{Helstrom1969, Holevo1973},
utilizing the Helstrom bound to saturate the BEP,
to characterize the performance of quantum demodulation protocol.
This differentiates our study from previous studies, which use quantum parameter estimation theory to bound achievable precision~\cite{Braunstein_1994, paris2009quantumestimationquantumtechnology, Giovannetti_2004, Giovannetti_2011, Escher_2011, Chaves_2013}.
We develop a quantum sensor protocol and compare its performance to
a classical electrically-small antenna of comparable active size.
We numerically compare
the performance of a  classical antenna to an ensemble of NV centers in diamond and identify the necessary defect densities for the quantum sensor to outperform the classical limits.
Finally, we also investigate the impact of sensor noise and imperfections in the applied quantum control to the performance of the demodulation algorithms, as well as develop a classical feedback control loop to mitigate the effects of such imperfections. 

The content is organized as follows. \sectionautorefname{\ref{sec:background}} sets up the study with preliminaries on signal processing and quantum state discrimination. 
\sectionautorefname{\ref{sec:quantum_model}} introduces the model Hamiltonian and error dynamics. 
\sectionautorefname{\ref{sec:quantum_protocol}} presents the main result of this work, the decoding protocol, as well as theory to calculate the optimal measurements and BEP. 
In Sec. \ref{sec:channel_capacity}, we calculate the Holevo information and define the channel capacity of the quantum receiver.
With \sectionautorefname{\ref{sec:comparison}, we use the Chu-limited channel capacity to compare an electrically-small classical receiver to the quantum protocols developed in this study.
\sectionautorefname{\ref{sec:error_suppression}} covers a discussion of error suppression to extend the sensing protocol.
\sectionautorefname{\ref{sec:conclusion}} concludes the work with a summary of our findings, an outlook on the practicality of our methods, and some future research directions.

\section{Background\label{sec:background}}

\subsection{Classical Communications}\label{subsec:classical-sensing-intro}

Details on the following review of classical communications can be found in standard texts such as \cite{Stern2004,Haykin2001}.
Wireless communication typically uses modulation techniques to encode low-frequency signals in high frequency (typically radio frequency) EM carrier waves. There are a wide variety of modulation schemes including amplitude modulation~(AM), frequency modulation~(FM), and phase modulation~(PM).
Phase-shift keying (PSK), which is a kind of PM, encodes digital information in the phase of the carrier.  We define an arbitrary PM signal as
\begin{equation}
    \label{eqn:psk_signal}
    S(t) = \sqrt{2P_{\rm{s}}} \cos( 2\pi f_{\rm s}t + \phi(t)) ,
\end{equation}
where $P_{\rm{s}}$ is the average transmitted power of the signal, 
$f_{\rm s}$ is the carrier frequency, and $\phi(t)$ is the carrier phase.
The time-dependent phase, $\phi(t)$, takes on discrete values known as \emph{symbols}, $\phi(t) \in \mathcal{S}$, where $\mathcal{S}$ is a finite set of phases used to encode data.  For example, the binary phase-shift keying (BPSK) modulation scheme has $\mathcal{S}\equiv\{0,\pi\}$, while quadrature phase-shift keying (QPSK) has $\mathcal{S}\equiv \{0,\pi/2,\pi,3\pi/2\}$.
In general, $l$-PSK encodes $l=|S|$ symbols.
The rate at which the phase changes is $f_{\rm{sym}}=1/T_{\rm{sym}}$ (\emph{symbol rate} or \emph{baud rate}). 
The rate of information (in terms of bits) encoded in the signal is given by the \emph{bitrate}, $f_{\rm b} =f_{\rm{sym}} \log_2 |\mathcal{S}|$, where $|\mathcal{S}|$ denotes the cardinality of $\mathcal{S}$. For example, Wireless LAN conventions have  $f_{\rm s}=2.4\ {\rm GHz}$ with bitrates $f_{\rm b}>1\ {\rm Mbps}$~\cite{ieee_wifi}. 

The receive chain in a wireless communication system is typically used to describe a series of components for processing the received signal and converting it into a digital signal. The receiver, which in the classical communication setting is typically an antenna, is the key component in the receive chain and fundamentally limits achievable data rates.
One metric for the performance of a receiver is its quality factor $Q=f_{\rm r}/B$, where $f_{\rm r}$ is the resonant frequency of the receiver and $B$ is the bandwidth at which the antenna operates effectively. 
Typically, for achieving higher data rates, a smaller $Q$ factor is desirable as it corresponds to a larger bandwidth. The theoretical upper limit to the maximum achievable data rate is given by the Shannon-Hartley theorem which establishes the relationship between the bandwidth and the channel capacity (or maximum data rate). 
For electrically-small classical antennas, the \emph{Chu limit} imposes a fundamental lower bound on the achievable quality factor $Q$ and consequently an upper bound on the achievable bandwidth,
\begin{equation}
    Q \geq Q_{\rm{Chu}} = \frac{1}{(k a)^3} + \frac{1}{k a},\textrm{ and } B\leq \frac{f_{\rm s}}{Q_{\rm Chu}},\label{eq:Chulimit}
\end{equation}
where $a$ is the radius of the smallest sphere enclosing the antenna, and $ k=2\pi f_{\rm s}/c$ is the wavenumber of the carrier signal~\cite{Chu1948,Harrington1960}.  
In order to characterize the actual performance of the receive chain, one must also take into account the demodulation scheme used to recover the encoded digital signal.
For PSK signals, the phase demodulation typically proceeds in three steps: (i) carrier recovery through synchronization with a local oscillator, (ii) demodulation to baseband by mixing the synchronized carrier wave with the received signal and filtering to extract in-phase~(I) and quadrature-phase~(Q) components, and (iii) symbol detection from measured I-Q components. The symbol time, $T_{\rm sym}=1/f_{\rm sym}$ provides an upper limit on the available time for performing these three steps for accurate real-time decoding. 

Information capacity sets the maximum bits (or symbols) that can be transmitted through a channel. 
By multiplying the information capacity by the symbol rate, we obtain its channel capacity---the maximum data rate in bits per second.
%
By the Shannon--Hartley theorem~\cite{Shannon1948}, the channel capacity in (bit/s) over an Additive White Gaussian Noise ~(AWGN) channel is 
\begin{align}
    \label{eqn:shannon_hartley}
    C = B\log_2\Big( 1 + \frac{P_{\rm{s}}}{N_0 B} \Big)
\end{align}
where $N_0/2$ is the noise power spectral density (PSD).
Using the Chu limit [Eq.~\eqref{eq:Chulimit}], we can create an upper-bound on the channel capacity for an electrically-small receiver,
\begin{align}
    \label{eqn:classical_capacity_limit}
    C \leq \frac{f_{\rm{s}}}{Q_{\mathrm{Chu}}}\log_2 \Big( 1 + \frac{P_{\rm{s}} Q_{\mathrm{Chu}}}{N_0 f_{\rm{s}}} \Big) 
\end{align}
which characterizes the maximum channel capacity (bit/s) by the carrier frequency $f_{\rm{s}}$, average transmitted power $P_{\rm{s}}$, and receiver size $a$ (through $Q_{\mathrm{chu}}$).

\subsection{Quantum State Discrimination}\label{subsec:quantum-sensing-intro}

Quantum state discrimination (QSD) is important in the context of a quantum receive chain since 
the received information is encoded in quantum states,
as we detail in Sec. \ref{sec:quantum_model}. 
Binary QSD refers to the task of distinguishing between two quantum states, $\hat\rho_0$ and $\hat{\rho_1}$, through  measurements on a target quantum system. A binary QSD protocol consists of a two-outcome measurement, $\{0,1\}$, given by the Positive Operator-Valued Measure (POVM) $\{\hat E_0,\hat E_1\}$, where outcome `0' is more likely if the state is $\hat\rho_0$ and outcome `1' is more likely if the state is $\hat\rho_1$. The optimal POVM minimizes the error probability, which is the probability of measuring outcome `0' when the system is in $\hat\rho_1$ or measuring `1' when the system is in $\hat\rho_0$. In absence of prior information regarding the state of the system, we assume that states $\hat \rho_0$ and $\hat \rho_1$ are equally likely, $P(\hat \rho_0)=P(\hat \rho_1)=1/2$. For $i,j\in\{0,1\}$, let $P(i|\hat \rho_j)=\text{Tr}[\hat E_i \hat \rho_j]$ denote the conditional probability of measuring outcome $i$ when the state is $\hat \rho_j$. The single-shot BEP is then given by 
\begin{align}
    \label{eqn:helstrom_bound}
    P_{\rm e, quantum} &= P(0|\hat \rho_1)P(\hat \rho_1)+ P(1|\hat \rho_0)P(\hat \rho_0),\\
    &=\frac{1}{2}\left(\Tr\left[\hat E_0\hat\rho_1\right]+\Tr\left[\hat E_1\hat\rho_0\right] \right)\\
    & =\frac{1}{2} -\frac{1}{2} \Tr\left[\hat E_0\big(\hat \rho_0-\hat \rho_1\big)\right] \\
    &= \frac{1}{2} -\frac{1}{2} \Tr\left[-\hat E_1\big(\hat \rho_0-\hat \rho_1\big)\right],
\end{align}
where the final two lines follow from the normalization of the POVM, i.e. $\hat E_0+\hat E_1=\mathbbm{1}$.
The \emph{Helstrom bound}~\cite{Helstrom1969} provides a lower limit to the single-shot BEP,
\begin{equation}
    P_{\rm e, quantum} \geq \frac{1}{2} - \frac{1}{2}||\hat \rho_0-\hat \rho_1||_1,
\end{equation}
where $||A||_1\equiv \text{Tr}[\sqrt{A^\dag A}]$ is the trace norm. The Helstrom bound is saturated when $\hat E_0$ is the projector onto the positive eigenspace of $\hat \rho_0-\hat \rho_1$. 
If we define observable $\mathcal{O} = \hat E_0 - \hat E_1$, then the BEP can be expressed as the difference of expectation values,
\begin{equation}
    \label{eqn:helstrom_bound_obs}
    P_{\rm e, quantum} = \frac{1}{2} - \frac{1}{4}\Big( \Tr[\mathcal{O}\rho_0] - \Tr[\mathcal{O}\rho_1] \Big).
\end{equation}
Intuitively, the Helstrom bound connects the distinguishability of two quantum states to their trace distance.

\section{Model for  a Quantum Receiver\label{sec:quantum_model}}

In this section, we model the dynamics of a quantum receiver for detection of PSK signals in the presence of noise.
Fig.~\ref{fig:quantum_schema}~(a) introduces the communication system.
Specifically, we consider the scenario where the quantum receiver is an ensemble of qubits (two-level systems) and the PSK signal is an electromagnetic wave that couples to the qubits directly. 
The quantum demodulation protocol or ``receive chain", which is depicted schematically in Fig.~\ref{fig:quantum_schema}~(b), involves the following steps: (i) the quantum receiver is prepared in a fiducial `sensing' state; (ii) the quantum receiver evolves under the signal; (iii) the quantum receiver is measured; and (iv) the transmitted bitstring is recovered through the measurement outcomes. 
These steps must happen within a symbol time $T_{\rm{sym}}$, while leaving time $T_{\rm{SPAM}}$ for state preparation and measurement (SPAM), see Fig.~\ref{fig:quantum_schema}~(c). 
We aim to quantify the performance of this quantum receive chain in the presence of realistic noise and errors.

\begin{figure}
    \includegraphics[width=\linewidth]{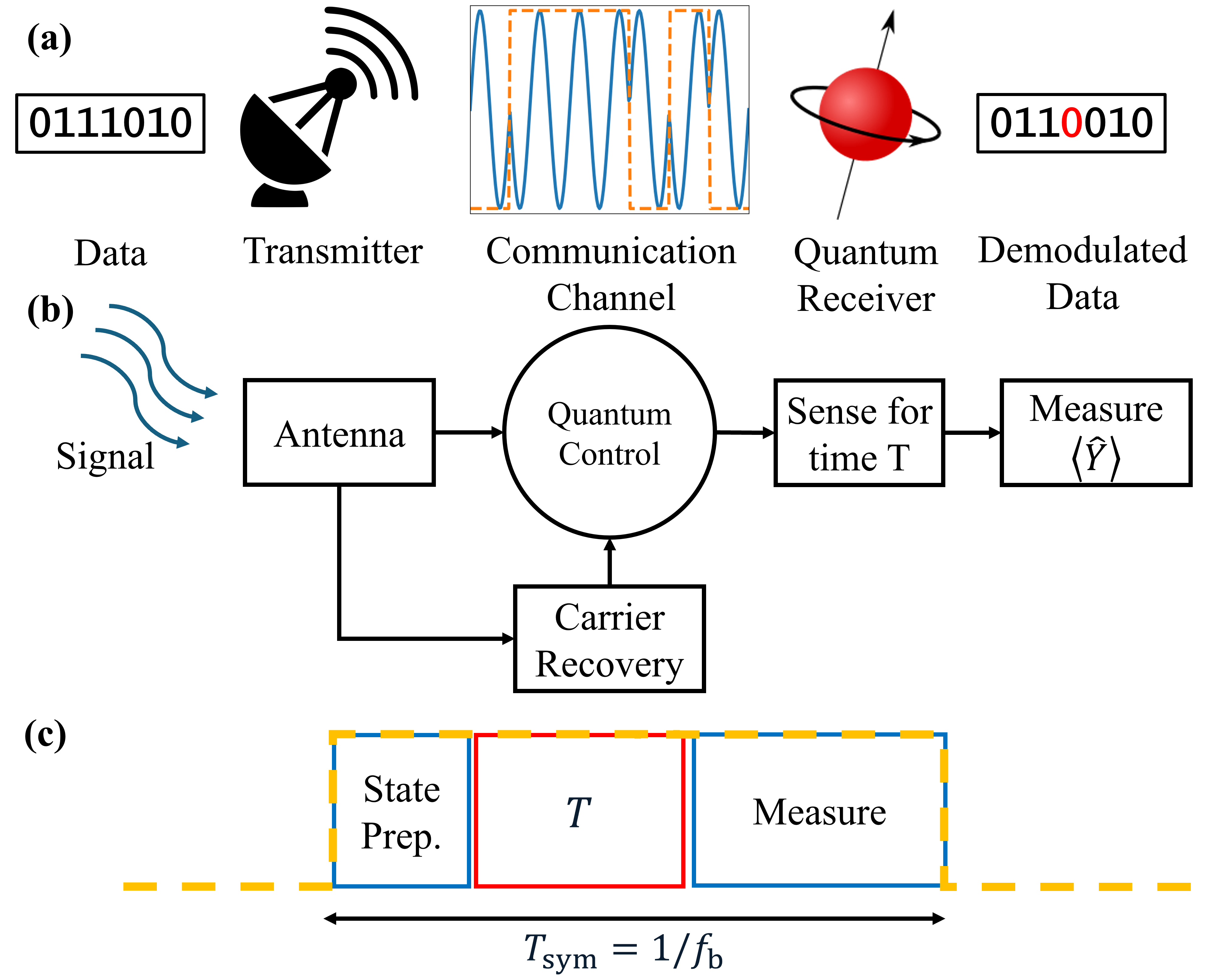}
    \caption{A quantum-enabled communications scheme. (a)~Schematic of a communication system involving a quantum receiver. The transmitter and channel components are classical, while the receiver is quantum. (b) Schematic for a signal demodulation protocol for PSK signals using a quantum receiver. The approach involves synchronizing a quantum sensor to the carrier wave, evolution and measurement of the sensor state; all of these steps must be accomplished within a symbol time.
    (c) Schematic of a single symbol discrimination: the qubit is prepared in the initial state, the qubit evolves in the presence of the signal, and then it is measured, determining the encoded symbol. Since the protocol utilizes discrimination per symbol, the above must occur within one symbol time, $T_{\textrm{sym}}$.
    }
    \label{fig:quantum_schema}
\end{figure}

Consider a quantum system of $n$ qubits with Hilbert space $\mathcal{H}$.
The dynamics of the quantum receiver under the effects of the signal, control, and noise are described in the laboratory frame by the Hamiltonian,
\begin{equation}
    \label{eqn:general_model}
    H(t) = \sum^n_{i=1} \Bigg( \underbrace{\frac{\omega_{\rm q}}{2}\sigma^z_i}_{\rm Splitting} + \underbrace{\vec\Omega(t)\cdot\vec\sigma_i}_{\rm Control} + \underbrace{\vec{s}(t) \cdot \vec\sigma_i}_{\textrm{Noisy Signal}} + \underbrace{\vec\eta_i(t)\cdot\vec\sigma_i}_{\rm Noise} \Bigg),
\end{equation}
where $\vec{\sigma}_i\equiv (\sigma^x_i,\sigma^y_i,\sigma^z_i)$ are the Pauli operators on the $i$th qubit, $\omega_{\rm q}$ is the qubit energy splitting, and throughout this work, we have chosen $\hbar=1$. 
The evolution occurs for a sensing time $T \leq T_{\rm{sym}}-T_{\rm{SPAM}}$, that is, within each symbol window, and leaving time for measuring and resetting.
This protocol fails if state preparation and measurement take longer than the time allotted to each symbol.

\emph{Control}---During the sensing phase, in which the qubits interact with the signal and noise, we consider control consisting of identical single-qubit rotations that can be applied along multiple axes, as specified by the control waveform vector $\vec{\Omega}(t)\equiv (\Omega^x(t),\Omega^y(t),\Omega^z(t))$.   
For our purposes, we consider control generated by resonant driving fields along $\sigma_i^x$, such that $\Omega^x(t)=\Omega^x \cos(\omega_q t)$, and static drives along $\sigma_i^z$, producing $\Omega^z(t)=\Omega^z/2$. The static drive can be realized, for example, through frame transformations or, in the case of atomic systems, via  ac-Stark shifts.
While the  single-qubit rotations applied during the sensing phase are non-entangling, we allow for the possibility that the initial state of the quantum receiver is entangled. 

\emph{Signal}---We model the signal as a vector $\vec s(t)\equiv(s^x(t),s^y(t),s^z(t))$ that uniformly couples to all qubits. Specifically, we consider two different modes of operation given by the nature of the signal coupling: 
\begin{subequations}
    \begin{align}
        {\rm Transverse}:\quad \vec s_{\rm T}(t) &= \left(s(t),0,0\right), \\ 
       {\rm Longitudinal} :\quad \vec s_{\rm L}(t) &= \left(0,0,s(t)\right).  
    \end{align}
\end{subequations}
Here, $s(t)=\Omega_{\rm s}\cos (\omega_{\rm s}t + \phi(t))$ is proportional to the carrier wave signal defined in Eq.~\eqref{eqn:psk_signal} with angular frequency $\omega_s=2\pi f_{\rm s}$. 
Longitudinal sensing is tuned to low frequencies, $\lesssim 10$ MHz, whereas transverse sensing is tuned to high frequencies around the natural qubit frequency $\omega_{\rm q}$, since the control generates a rotating frame that transforms the signal differently along the x and z axes.
The rotating frame transform is discussed more in the following section.
The transmitted bitstring is encoded by PSK modulation of the time-dependent phase $\phi(t)$. 
The amplitude of the carrier/qubit interaction $\Omega_{\rm s}\propto \sqrt{P_{\rm{s}}}$ depends on the particularities of the quantum sensing platform.

\emph{Signal and Noise}---We model two noise processes---phase noise in the signal and additive noise. 
Phase noise may arise from noisy transmission of the signal through the communication medium or phase mismatch between the local oscillator and the signal. 
We take $\overline{\cdots}$ to denote the ensemble average over realizations of the noise processes.
Phase noise $\varphi(t)$ enters the signal via
\begin{equation}
    \label{eqn:signal}
     s(t) = \Omega_{\rm{s}}\cos(\omega_{\rm{s}} t+\phi(t)+\varphi(t)).
\end{equation}
The phase noise is a stationary Gaussian process, $\varphi(t) \sim \mathcal{N}(m, \sigma^2)$, where $m$ is the mean phase and $\sigma^2$ the variance. Its auto-covariance is $\overline{\Delta \varphi(t) \Delta\varphi(0)} = \sigma^2 \delta(t)$, where $\Delta\varphi(t)=\varphi(t)- m$.
We model the additive noise  $\vec{\eta}_i(t)\cdot\vec{\sigma_i}$ semi-classically with Gaussian multi-axis stationary zero-mean processes, 
$\vec\eta_i(t)\equiv(\eta_i^x(t),\eta_i^y(t),\eta_i^z(t))$. The correlation functions $C^{\mu\nu}_{ij}(t)=\delta_{\mu\nu} \overline{\eta^\mu_i(t)\eta^\nu_j(0)}$, $\mu,\nu \in \{x,y,z\}$ quantify the degree of spatiotemporal correlation between the noise acting on the $i^{\rm th}$ and $j^{\rm th}$ qubits. Spatiotemporally correlated models have been used to model noise in superconducting \cite{Bylander_2011, Krantz_2019, Wilen_2021}, semiconductor \cite{Connors_2022}, and trapped ion \cite{Frey_2020} quantum devices.  
We assume that the signal and noise are uncorrelated, so that $\overline{s^\mu(t_1)\eta_i^\nu(t_2)}=0$ at all times $t_1,\,t_2$.
The frequency-domain effects of the noise are captured by the noise power spectral densities (PSD), which are defined as the Fourier transform of the noise correlation functions, $S^{\gamma\delta}_{ij}(\omega) = \int^T_0 C^{\gamma\delta}_{ij}(t)e^{-i\omega t}dt$. 
Then the total noise power is $P_N = \int^\infty_{-\infty} \frac{d\omega}{2\pi}\sum_{\gamma,i}S^{\gamma\gamma}_{ii}(\omega)$, and we define the signal-to-noise ratio as $\textrm{SNR}=P_{\rm{s}}/P_N$.

\subsection{Rotating Frame Dynamics \label{ssec:frame_transforms}}

In the rotating frame, the Hamiltonian of Eq.~\eqref{eqn:general_model} becomes
\begin{align}
 H^{(\textrm{rf})}(t)= \sum^n_{i=1}\Big[ H_{c,i}(t) + U_{\textrm{q},i}^\dag(t) \big( 
    \vec{s} \cdot \vec\sigma_i 
    + \vec\eta_i\cdot\vec\sigma_i
    \big) U_{\textrm{q},i}(t) \Big],   
\end{align}
 where $U_{\textrm{q},i}(t)=e^{-i\omega_{\textrm{q}} t\sigma^z_i/2}$, and 
the rotating-frame control Hamiltonian is   
\begin{align}
H_{c,i}(t)=\Omega^x \sigma^x_i /2 + \Omega^z \sigma^z_i /2.
\end{align}
The rotating-frame control Hamiltonian follows from the rotating-wave approximation (RWA), which holds in the limit $\omega_q T_{\rm{sym}} \gg 1$.

To separate the contributions of the noise and signal from the control, we transform into the ``toggling frame" or interaction picture associated with the rotating-frame control Hamiltonian. In this frame, the receiver dynamics are generated by the toggling-frame Hamiltonian,
\begin{align}\label{eqn:toggling_frame}
    \tilde H(t) = 
    \sum^n_{i=1} U_{c,i}^\dag(t) U_{\textrm{q},i}^\dag(t) \Big( \vec s\cdot\vec\sigma_i + \vec\eta_i\cdot\vec\sigma_i \Big) U_{\textrm{q},i}(t) U_{c,i}(t),
\end{align}
where $U_{\textrm{c},i}(t)=\mathcal{T}_+\exp[-i\int^t_0 H_{\textrm{c},i}(s)ds]$ is the control propagator on qubit $i$. Evolution of the receiver in the laboratory and toggling frames is related by $U(t)=U_{\textrm{q}}(t)U_{\textrm{c}}(t)\tilde{U}(t)$, where $U(t)=\mathcal{T}_+\exp[-i\int^t_0 H(s)ds]$ is the laboratory-frame propagator generated by the Hamiltonian in Eq.~\eqref{eqn:general_model},  $\tilde{U}(t)=\mathcal{T}_+\exp[-i\int^t_0 \tilde{H}(s)ds]$ is the toggling-frame propagator, $U_{\textrm{q}}(t)=\bigotimes_{i=1}^n U_{\textrm{q},i}(t)$, and $U_{\textrm{c}}(t)=\bigotimes_{i=1}^n U_{\textrm{c},i}(t)$.

We can write the toggling-frame Hamiltonian in a more compact form by introducing the control matrix on qubit $i$, $\mathbf{R}_i(t)$ \cite{Green_2013}. For $\mu,\nu\in\{x,y,z\}$, the control matrix is a $3\times 3$ matrix with elements
\begin{equation}
    \label{eqn:switching_function}
    \mathbf{R}^{\mu\nu}_i(t)=\Tr\big[ U_{\textrm{c},i}^\dag(t)U_{\textrm{q},i}^\dag(t)\sigma^\mu_i U_{\textrm{q},i}(t)U_{\textrm{c},i}(t) \sigma^\nu_i \big]/2^n. 
\end{equation}
In terms of the control matrix, the toggling-frame Hamiltonian then becomes
\begin{equation}
    \label{eqn:general_Ham_toggling}
    \tilde H(t) = \sum_{i=1}^n\sum_{\mu\nu\in\{x,y,z\}}\Big[ s^\mu(t) + \eta^\mu_i(t) \Big]\mathbf{R}^{\mu\nu}_i(t)\sigma^\nu_i.
\end{equation}
From this expression, we see that $\mathbf{R}^{\mu\nu}_i(t)$ describes how control modifies the response of the receiver to the signal and noise in the time domain.

\subsection{Modeling Noisy Dynamics: Cumulant Expansion\label{ssec:cumulant_expansion}}
Here, we model the noisy dynamics of the quantum receiver using time-dependent perturbation theory and a cumulant expansion~\cite{kubo_1962,Paz_Silva_2017, zhou2022quantum,quiroz2021quantifying}.
Since the noise acting on the receiver is a stochastic process, 
the expected value of an observable is computed  using an ensemble average over noise trajectories.
For an initial state of the receiver, $\rho_{\rm in}$, the expected value of an invertible observable $\mathcal{O}$ at time $T$ in the toggling frame is given by
\begin{align}
    \label{eqn:avg_obs}
    \overline{\braket{\mathcal{O}(T)}}&=\mathrm{Tr}\left[ \Lambda_{\mathcal{O}}(T) \rho_{\rm in}\mathcal{O}\right]  \\
    \textrm{with } \Lambda_\mathcal{O}(T)&= \overline{\mathcal{O}^{-1} \tilde U^\dagger(T) \mathcal{O} \tilde U(T)}, 
\end{align}
where $\overline{\cdots}\,$ denotes the ensemble average over noise realizations, and $\braket{\cdots}$ denotes the quantum mechanical expectation value with respect to the state of the receiver~\cite{Paz_Silva_2017}. The toggling-frame expectation value above is equivalent to the lab-frame expectation value when
$U_{\textrm{q}}(T)U_{\textrm{c}}(T)=I$. Although the control used in our quantum receive chain does not generally generate a net identity, we can enforce the condition $U_{\textrm{q}}(T)U_{\textrm{c}}(T)=I$ by applying fast qubit rotations prior to measurement, making $\overline{\braket{\mathcal{O}(T)}}$ equivalent to the lab-frame expectation value.

In the weak-signal and  noise regime ($|\vec\eta |T,|\vec s |T\ll 1$), we can evaluate the effective error operator $\Lambda_{\mathcal{O}}(T)$ using a perturbative cumulant expansion,
\begin{align}
    \label{eqn:error_operator}
    \Lambda_\mathcal{O}(T) &= e^{\mathcal{C}_{\mathcal{O}}(T)}.
\end{align}
And because the signal and noise are uncorrelated, the cumulant expansion separates into signal and noise-dependent terms, $\mathcal{C}_{\mathcal{O}}(T) = \mathcal{C}_{\mathcal{O},s}(T)+\mathcal{C}_{\mathcal{O},n}(T)$, where
\begin{align}
&\mathcal{C}_{\mathcal{O},s}(T)=\sum^\infty_{k=1} (-i)^k \frac{\mathcal{C}^{(k)}_{\mathcal{O},s}(T)}{k!},\\
&\mathcal{C}_{\mathcal{O},n}(T)=\sum^\infty_{k=1} (-i)^k \frac{\mathcal{C}^{(k)}_{\mathcal{O},n}(T)}{k!},
\end{align}
and $\mathcal{C}^{(k)}_{\mathcal{O},s}(T)$ $[\mathcal{C}^{(k)}_{\mathcal{O},n}(T)]$ denotes the cumulant of order $k$ on the signal [noise]. 
We consider a zero-mean noise model ($\overline{\eta_i(t)}=0$) so $\mathcal{C}^{(1)}_{\mathcal{O},n}(T)=0$.
We assume a weak-signal/noise regime, so that the signal fluctuations, as captured by the second cumulant, are negligible, and that $|\mathcal{C}^{(2)}_{\mathcal{O},s}(T)|/|\mathcal{C}^{(2)}_{\mathcal{O},n}(T)|\ll 1$, i.e., a small variance in the phase error.
Then the dynamics are well approximated by truncating the expansion up to leading non-zero order,
\begin{align}
    \mathcal{C}_{\mathcal{O},s}(T)\approx -i \mathcal{C}_{\mathcal{O},s}^{(1)}(T), \\
    \mathcal{C}_{\mathcal{O},n}(T)\approx -\mathcal{C}_{\mathcal{O},n}^{(2)}(T)/2.
\end{align}
Here, the first cumulant of the signal is
\begin{align}
    \mathcal{C}_{\mathcal{O},s}^{(1)}(T) = \sum_{i=1}^n\sum_{\mu\nu\in\{x,y,z\}}\int^T_0 dt\ \overline{s^\mu(t)} \mathbf{R}^{\mu\nu}_i(t) \nonumber\\ 
    \times
    \Big[ \sigma^\nu_i - \mathcal{O}^{-1}\sigma^\nu_i\mathcal{O} \Big],
\end{align}
where the signal is stochastic in the phase noise.
The signal mean is evaluated
\begin{align}
    \overline{s^\mu(t)} = \Omega_s e^{-\sigma^2/2} \cos(\omega_s t+\phi(t)+\mu),
\end{align}
for $\mu=x$ in the transverse case, $\mu=z$ in the longitudinal case, and $\overline{s^\mu(t)}=0$ otherwise.
Because the additive noise is zero-mean, its leading order contribution is the second cumulant, 
\begin{align}
    \frac{1}{2} \mathcal{C}^{(2)}_{\mathcal{O},n}(T) &= \sum^N_{i,j=1}\sum_{ \substack{\mu,\nu\ \in\\ \{x,y,z\}}} \Big[ \chi^{\mu\nu}_{ij}(T) \left[\mathcal{A}_{\mathcal{O}}\right]^{\mu\nu}_{ij} 
    + \psi^{\mu\nu}_{ij}(T) \left[\mathcal{B}_{\mathcal{O}}\right]^{\mu\nu}_{ij} \Big] 
    \label{eqn:general-2nd-cumulant-1},
\end{align}
where 
\begin{align}
    [\mathcal{A}_{\mathcal{O}}]_{ij}^{\mu\nu} =
    \sigma^\mu_i\sigma^\nu_j - \sigma^\mu_i O^{-1}\sigma^\nu_j O  \nonumber\\
    - O^{-1}\sigma^\mu_i O \sigma^\nu_j + O^{-1} \sigma^\mu_i\sigma^\nu_j O 
    \label{eqn:general-2nd-cumulant-2} \\
    \textrm{and } [\mathcal{B}_{\mathcal{O}}]_{ij}^{\mu\nu} = O^{-1}[\sigma^\mu_i,\sigma^\nu_j]O - [\sigma^\mu_i,\sigma^\nu_j] 
    \label{eqn:general-2nd-cumulant-3}
\end{align} 
are observable-dependent operators, while
\begin{align}
    \chi^{\mu\nu}_{ij}(T) &= \sum_{\substack{\gamma,\delta \in \\ \{x,y,z\}}}\int^\infty_0 \frac{d\omega}{2\pi}\Re\Big[ S^{\gamma\delta}_{ij}(\omega) \mathcal{F}^{\gamma\mu\delta\nu}_{ij}(\omega,T) 
    \label{eqn:general-2nd-cumulant-4} \Big] \end{align}
    represents noise-induced decay, and
    \begin{align}
    \psi^{\mu\nu}_{ij}(T) &=  \frac{1}{2} \sum_{\substack{\gamma,\delta \in \\ \{x,y,z\}}}\int^\infty_{-\infty} \frac{d\omega}{2\pi}S^{\gamma\delta}_{ij}(\omega) \mathcal{G}^{\gamma\mu\delta\nu}_{ij}(\omega,T)
    \label{eqn:general-2nd-cumulant-5}
\end{align}
can generate both noise-induced decay and coherent error~\cite{cerfontaine_2021_prr, Watkins_2025}. The noise-induced decay rates depend on the frequency-domain overlap between a noise PSD and a control filter function (FF): $\mathcal{F}^{\gamma\mu\delta\nu}_{ij}(\omega,T) = F^{\gamma\mu}_i(\omega,T)F^{\delta\nu}_j(-\omega,T)$ or $\mathcal{G}^{\gamma\mu\delta\nu}_{ij}(\omega,T) = \int^T_0 dt_1 e^{i\omega t_1}R^{\gamma\mu}(t_1) F^{\delta\nu}_j(\omega,t_1)$, with $F^{\gamma\mu}_i(\omega,T) = \int^T_0 dt e^{-i\omega t}R^{\gamma\mu}_i(t)$. 
In later sections (see for e.g., Sec.~\ref{subsec:opt-meas-basis}), we apply this model to quantify noise on receiver performance metrics, such as BEP.

\section{Quantum Demodulation Protocol\label{sec:quantum_protocol}}

The general demodulation protocol for a PSK signal using a quantum receiver is schematically outlined in \figureautorefname{\ref{fig:quantum_schema}}~(b). Analogous to a  classical PSK demodulation algorithm, the quantum protocol proceeds in the following steps for each symbol encoded in the PSK signal:
\begin{enumerate}
    \item The quantum state of the receiver is initialized/reset to the desired sensing state, $\hat \rho_{\rm in}$.
    \item The applied quantum control drive is synchronized classically to the signal carrier~\cite{schultz_2022}
    \item The quantum state evolves in the presence of the signal and control for predetermined amount of sensing time, $T$ to a state $\hat\rho_\phi (T)$. The control is chosen to maximally distinguish the different possible discrete symbols represented by $\phi$.
    \item The state is measured in an appropriate basis $\{\hat E_\xi\}$ corresponding to the different symbols $\xi \in \mathcal{S}$. The basis is chosen to saturate the Helstrom bound that minimizes the BEP.
\end{enumerate}
While the majority of this work focuses on BPSK demodulation, we present a generalization in \appendixautorefname{\ref{app:quantum_psk}} to demodulate any PSK signal, such as quadrature phase-shift keying (QPSK) and quadrature amplitude modulation (QAM), with our quantum protocol.
An important constraint on the demodulation protocol requires that all steps, i.e., state preparation, evolution and measurement and reset, are completed within a symbol time, $T_{\rm sym}$. Thus, given a quantum sensor system, its gate times and measurement/reset times can be used to estimate maximum achievable data transfer rates.

\subsection{Control Scheme}\label{subsec:control scheme} 

We will now delve into the control scheme for the demodulation protocol of a transverse signal coupling in the quantum receiver model introduced in Sec.~{\ref{sec:quantum_model}}.
The following Hamiltonians generate the interaction picture evolution operator, $U_c(t)$, that is used in the cumulant expansion of Sec.~{\ref{ssec:cumulant_expansion}}.
Because of the similarity between $\tilde{H}_{\rm T}$~[Eq.~\eqref{eqn:transversal}] and $\tilde H_{\rm L}(t)$~[Eq.~\eqref{eqn:longitudinal}], with the substitution $x\leftrightarrow z$, we restrict our analysis to the transverse coupling Hamiltonian with the knowledge that the results straightforwardly generalize to the longitudinal couplings. 
The longitudinal case is outlined in Appendix~{\ref{subsubsec:control for longitudinal}}.

\subsubsection{Transverse coupling}\label{subsec:control for transverse} 
A transverse coupling between the signal and the receiver is advantageous when the natural frequency of the qubit is close to the signal carrier frequency, $\omega_{\rm q}\approx \omega_{\rm s}$. By choosing a constant longitudinal control (i.e., along the $z$ axis), $\vec\Omega(t) \equiv\left(0,0,\Omega_{\rm c}/2\right)$, we get the lab-frame Hamiltonian
\begin{equation}
    \label{eqn:transverse_lab_frame}
    H_{\rm T}(t) = \sum^n_i \Big( \left(\frac{\omega_{\rm q}}{2}+\frac{\Omega_{\rm c}}{2}\right)\sigma^z_i +s(t)\sigma^x_i + \vec\eta_i(t)\cdot\vec\sigma_i \Big),
\end{equation} 
where $s(t)$ is the signal with phase noise [Eq.~\eqref{eqn:signal}].
As discussed in Sec.~{\ref{ssec:frame_transforms}}, we go from the laboratory frame to the rotating frame, and then from the rotating frame to the toggling frame.
We apply Eq.~\eqref{eqn:toggling_frame} to transverse coupling, meaning $U_c(t) U_q(t) = \exp \Big[-i\big(\omega_{\rm q} + \Omega_{\rm c}\big)t\sigma^z_i/2\Big]$.
Therefore, the relevant control matrix elements for the signal are
$R^{xx}_i(t) = \cos(\omega_{\rm q}+\Omega_c)t$
and
$R^{xy}_i(t) = -\sin(\omega_{\rm q}+\Omega_c)t$,
so that
the Hamiltonian in the interaction picture,
with some trigonometric multiplicative identities, becomes
\begin{widetext}
\begin{equation}
    \begin{split}
    \tilde H_{\rm T}(t) = \sum^n_i
    \Bigg( \frac{\Omega_{\rm s}}{2} \Big( 
    \Big[ \cos\big((\omega_s+\omega_q+\Omega_c)t+\phi(t)+\varphi(t) \big) 
    +\cos\big((\omega_s-\omega_{\rm q}-\Omega_c)t+\phi(t)+\varphi(t)\big) \Big]\sigma^x_i \\
    - \Big[ \sin\big((\omega_s+\omega_{\rm q}+\Omega_c)t+\phi(t)+\varphi(t)\big) 
    +\sin\big((-\omega_s+\omega_{\rm q}+\Omega_c)t-\phi(t)-\varphi(t)\big) \Big]\sigma^y_i 
    +  \sum_{\mu\nu\in\{x,y,z\}}\eta^\mu_i(t) R^{\mu\nu}_i(t)\sigma^\nu_i \Bigg).
    \end{split}  
\end{equation}
\end{widetext}
By choosing $\Omega_{\rm c} = \omega_{\rm s} - \omega_{\rm q}$, the sensor is resonant with the signal,
so under the rotating-wave approximation (RWA), for $\omega_s T_{\rm{sym}} >> 1$, one gets
\begin{equation}
    \label{eqn:transversal}
    \begin{split}
      \tilde H_{\rm T}(t) = \\
      \sum^n_i\Bigg( \frac{\Omega_{\rm s}}{2}\Big( \cos\big(\phi(t)+\varphi(t)\big) \sigma^x_i + \sin\big(\phi(t)+\varphi(t)\big)\sigma^y_i \Big) \\
      +  \sum_{\mu\nu\in\{x,y,z\}}\eta^\mu_i(t) R^{\mu\nu}_i(t)\sigma^\nu_i \Bigg),
    \end{split}
\end{equation}
where the noise is transformed, as described in Sec.~{\ref{ssec:frame_transforms}}. 
In the toggling frame, it is clear that the qubit evolves under an effective Hamiltonian dependent on the encoded phase $\phi(t)$.

\begin{figure}[t]
    \centering
    \includegraphics[width=0.8\linewidth]{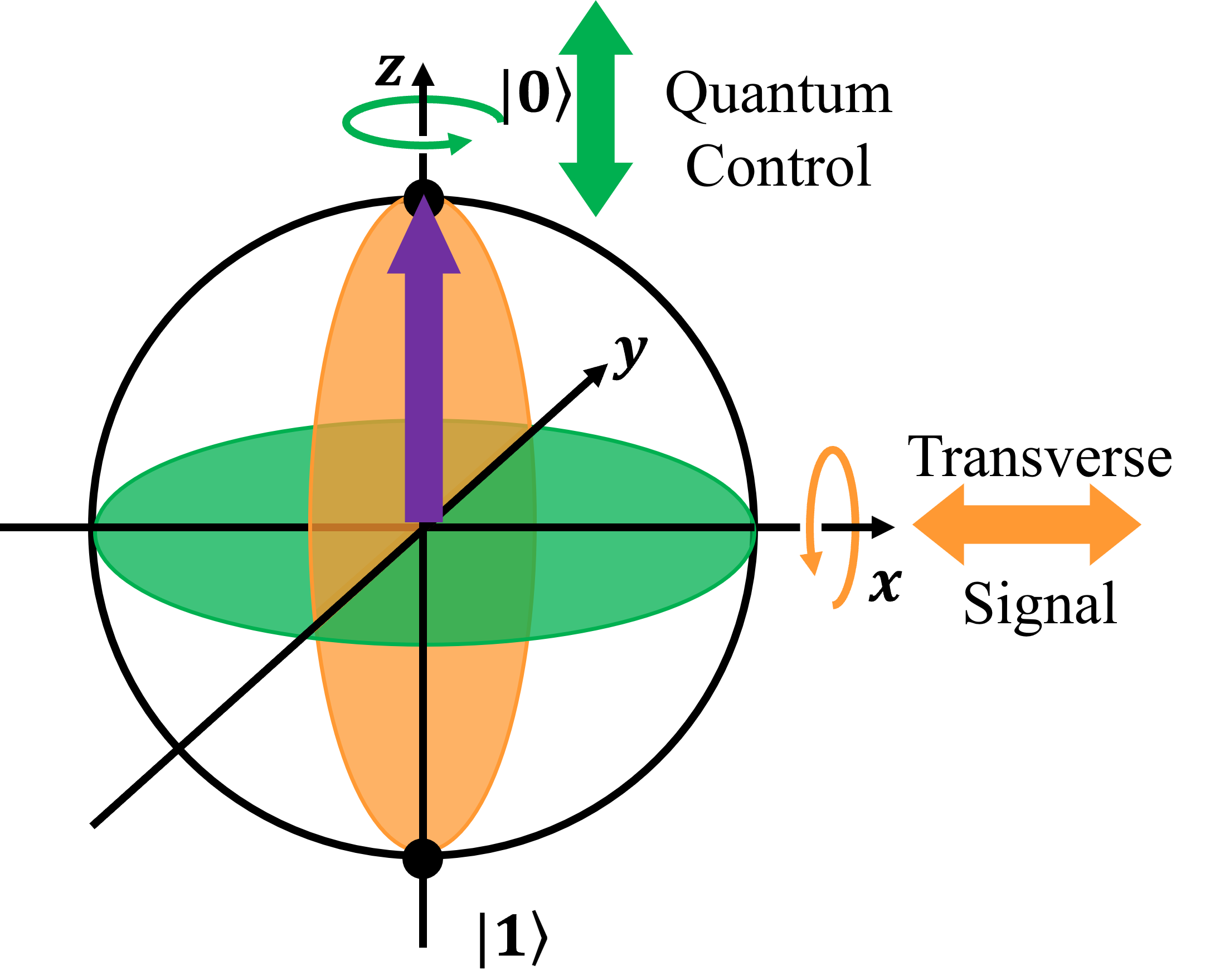}
    \caption{Schematic representation of a qubit in the quantum demodulation protocol.
    A sensor is prepared the ground state $\ket{0}^{\otimes n}$ and measured along the $y$-axis to demodulate the signal. The orange arrow denotes the direction of the signal coupling, the green arrow the control axis, and the purple arrow the initial state.
    Transverse coupling of the signal requires constant longitudinal control.
    }
    \label{fig:sensing_modes}
\end{figure}
}
\label{fig:bloch_spheres}

\subsection{Initial state\label{ssec:initial_states}}
The Hamiltonian for transverse coupling, $\tilde H_{\rm T}$, corresponds to a global single-qubit rotation around an axis in the $x$-$y$ plane; 
its Rabi frequency is proportional to the signal strength.
For sensing, we consider the following initial states: 
\begin{itemize}[leftmargin=*]
    \item {\bf Product states}: Single qubit states orthogonal to the $x$-$y$ plane (i.e $z$-basis states) are most sensitive to these global rotations. Thus, we choose the following product initial state of the ensemble,
\begin{align}
    \hat \rho_{\rm in}&=
        \ket{0}\bra{0}^{\otimes n}\label{eq:rho-prod}
\end{align}
\item {\bf Entangled states}: Entangled states can provide greater sensitivity to global rotations compared to product states. The Greenberger–Horne–Zeilinger (GHZ) state achieves the Heisenberg limit for sensing. Thus, we use the following GHZ state, 
\begin{align}
    \hat \rho_{\rm in}&=
        \ket{\psi^{+}_{\rm GHZ}}\bra{\psi^{+}_{\rm GHZ}} \label{eq:rho-ghz}
\end{align}
where $\ket{\psi_{\rm GHZ}^{\pm}}=\frac{1}{\sqrt{2}}\left(\ket{+}^{\otimes n}\pm\ket{-}^{\otimes n}\right)$ and $\ket{\pm}=\frac{1}{\sqrt{2}}(\ket{0}+\ket{1})$.
\end{itemize}
Note that for longitudinal coupling, we make the substitutions $\{\ket{0},\ket{1}\}\leftrightarrow \{\ket{+},\ket{-}\}$.

\subsection{Optimal Measurement and BEP}\label{subsec:opt-meas-basis}

The optimal measurement basis can be derived by saturating the Helstrom bound (see Sec.~{\ref{subsec:quantum-sensing-intro}}). For this discussion, we make the following simplifying assumptions: (i) we restrict to the scenario of BPSK signals, i.e. $\phi(t)=b\pi$ corresponding to binary symbols $b\in \{0,1\}$, and (ii) we assume no noise ($\vec \eta=\vec0$). We also restrict our discussion to the transverse signal coupling, noting again that the results can be straightforwardly generalized to the longitudinal case. With these constraints, the Hamiltonian evolution in the toggling frame simplifies to
\begin{align}
    \tilde H_{\rm T}[b] =(-1)^b\sum_{i=1}^n\frac{\Omega_{\rm s}}{2}\sigma^x_i,\ \textrm{with } b\in\{0,1\}. \label{eq:transverse-H}
\end{align}
The corresponding output state measured following a sensing time $T$, is $\rho_b(T)=U_0(T)\tilde U(T)\rho_{\rm in}\tilde U^\dagger(T) U_0^\dagger(T) = e^{-i\tilde H_{\rm T}[b]T}\hat \rho_{\rm in}e^{i\tilde H_{\rm T}[b]T}$, for measurements synchronized with the control, such that, $U_0(T)=1$. The optimal POVMs $\{\hat E_0,\hat E_1\}$ saturating the Helstrom bound are given by the positive and negative eigenspaces of the difference operator, $\hat \rho_1(T)-\hat \rho_0(T)$.

\subsubsection{Single-qubit Measurement\label{sssec:single_qb_meas}}

 First, let us consider a single-qubit sensor ($n=1$) in the initial state $\hat\rho_{\rm in}=\ket{0}\bra{0}$, then we have
\begin{align}
    \label{eqn:y_measurement}
    &\rho_b(T)  = \frac{1}{2}\Big(\hat I + \cos \Omega_{\rm{s}}T \hat\sigma^z + (-1)^b \sin \Omega_{\rm{s}}T \hat\sigma^y \Big),\\
    &\rho_1(T)-\rho_0(T)=-\sin\big(\Omega_{\rm{s}}T\big)\sigma^y.
\end{align}
The optimal basis (that saturates the Helstrom bound) for discriminating the two symbols is a measurement  in the $y$-basis, $\{\hat E_0,\hat E_1\}\equiv\{\ket{+i}\bra{+i},\ket{-i}\bra{-i}\}$ with $\ket{\pm i}=\frac{1}{\sqrt{2}}\left(\ket{0}\pm i\ket{1}\right)$. It follows from Eq. (\ref{eqn:helstrom_bound_obs}) that the Helstrom bound is saturated by measuring the observable $\sigma^y = \hat E_0 - \hat E_1$, so, in the absence of noise, that the optimal BEP is
\begin{align}
    P_{\rm e} &=\frac{1}{2}-\frac{1}{4}\left(\braket{\sigma^y}_1-\braket{\sigma^y}_0\right)=\frac{1}{2}\big(1-\sin(\Omega_{\rm{s}} T)\big),
\end{align}
where $\braket{\sigma^y}_b=\Tr[\rho_b \sigma^y]$, $b \in \{0,1\}$. 
From this expression, we see that $P_{\rm e}$ vanishes when the sensor evolves for the optimal sensing time  $T_{\rm{opt}} = \frac{\pi}{2\Omega_{\rm s}}$. We show the dependence of BEP as a function of time $T$ in Fig.~\ref{fig:bit_error_dephasing}~(a). 

Now, we determine the optimal measurement $\mathcal{O}$ in the scenario of noisy transverse sensing with additive Z-dephasing noise $\vec\eta(t) =(0,0,\eta(t))$ and signal phase noise $\varphi(t)$, which were introduced in Sec.~\ref{sec:quantum_model}.
We can use Eq.~{\eqref{eqn:avg_obs}} and the leading order cumulants to calculate the noise-averaged expectation value $\overline{\braket{\mathcal{O}}}$.
For the transverse signal with phase noise, the first cumulant is given by
\begin{align}
    -i\mathcal{C}^{(1)}_{\mathcal{O},{\rm{s}}}(T) =
    -i\frac{\Omega_{\rm s} T}{2} (-1)^b e^{-\sigma^2/2}\cos\mu \left( \sigma^x - \mathcal{O}^{-1} \sigma^x \mathcal{O} \right),
\end{align}
For additive Z-dephasing noise, the general expressions for the second cumulant in Eqs.~(\ref{eqn:general-2nd-cumulant-1}-\ref{eqn:general-2nd-cumulant-5}) simplify to a single decoherence parameter $\chi(T)$, corresponding to $\mu=\nu=\gamma=\delta=z$ in Eq.~\eqref{eqn:general-2nd-cumulant-4},
\begin{align}
    \chi(T) &= \int^\infty_0 \frac{d\omega}{2\pi} S(\omega)\Big( \frac{\sin(\omega T/2)}{\omega/2}\Big)^2. 
    \label{eqn:z_dephasing_rate}
\end{align}
So the second cumulant is then
\begin{align}
    \frac{1}{2} \mathcal{C}^{(2)}_{\mathcal{O},n}(T) = \chi(T) [\mathcal{A}_\mathcal{O}]^{zz}
\end{align}
and $[\mathcal{A}_\mathcal{O}]^{zz}=4I$ or $0$, depending on the observable $\mathcal{O}$.
We find that $\mathcal{O}\neq \sigma^x$, otherwise the first cumulant (encoding the phase information) would be zero.
Therefore the truncated cumulant series is
\begin{align}
    \mathcal{C}_{\mathcal{O}}(T) = -i (-1)^b e^{-\sigma^2/2}\cos\mu\,\Omega_{\rm s} T  \sigma^x - 2\chi(T)I
\end{align}
which can now be plugged into Eq.~\eqref{eqn:error_operator}.
Lets define the expectation $\braket{\mathcal{O}}_b = \Tr[\mathcal{O}\rho_b]$ for the two inputs $b=\{0,1\}$.
Then, the noise-averaged expected value of an observable $\mathcal{O}$ becomes
\begin{align}
    \overline{\braket{\mathcal{O}}_b} &= e^{-2\chi(T)} \Tr\Big[ e^{-i(-1)^b \theta \sigma^x} \rho_{\rm in} \mathcal{O}\Big] \nonumber\\
    &= e^{-2\chi(T)} \Tr\Big[ e^{-i(-1)^b \theta \sigma^x} \rho_{\rm in} e^{i(-1)^b \theta \sigma^x/2} \mathcal{O}\Big]
\end{align}
where we have abbreviated the angle $\theta =  e^{-\sigma^2/2}\cos\mu\, \Omega_s T$ and used the fact that $\mathcal{O}$ and $\sigma^x$ anti-commute.
It is clear that the dynamics are identical to the noiseless case, with the addition of exponential decay $e^{-\chi(T)}$ to the amplitude and $e^{-\sigma^2/2}$ to the frequency.
So as before, it is clear that $\mathcal{O}=\sigma^y$ is the optimal measurement.
As a result, we get the BEP in the presence of additive sensor noise and signal phase noise,
\begin{align}
    \label{eq:q_bep_single_noise_offset}
    P_{\rm e} = \frac{1}{2} - \frac{1}{2}e^{-2\chi(T)} \sin\left( \tilde\Omega_{\rm s} T \right),
\end{align}
where we define an effective Rabi frequency $\tilde\Omega_{\rm s} = e^{-\sigma^2/2}\cos\mu\,\Omega_{\rm s}$, as the signal strength modulated by the phase noise.
We note that the dephasing noise introduces an exponential damping term, and that the phase noise introduces an exponential damping in the argument of the sine function. Both of these lead to a larger BEP. 
Any noise in the phase of the signal can be accounted for in the calibration of the optimal time which shifts to $T_{\rm opt}=\pi/(2\tilde\Omega_{\rm s})$, in the absence of dephasing.
If the Z-dephasing noise is white, i.e. with a constant PSD $S(\omega)=\Gamma$, and phase noise is absent, the decoherence rate is $\chi(T)=\Gamma T/2$ 
, and the optimal sensing time becomes $T_{\rm opt} = \frac{1}{\tilde\Omega_{\rm s}}\tan^{-1}(\frac{\Omega_{\rm s}}{\Gamma})$. 
For more details on this derivation, see App.~{\ref{app:noisy_optimal_measurement}}. 

\subsubsection{Multi-qubit Measurement}

\begin{figure}
\includegraphics[width=\linewidth]{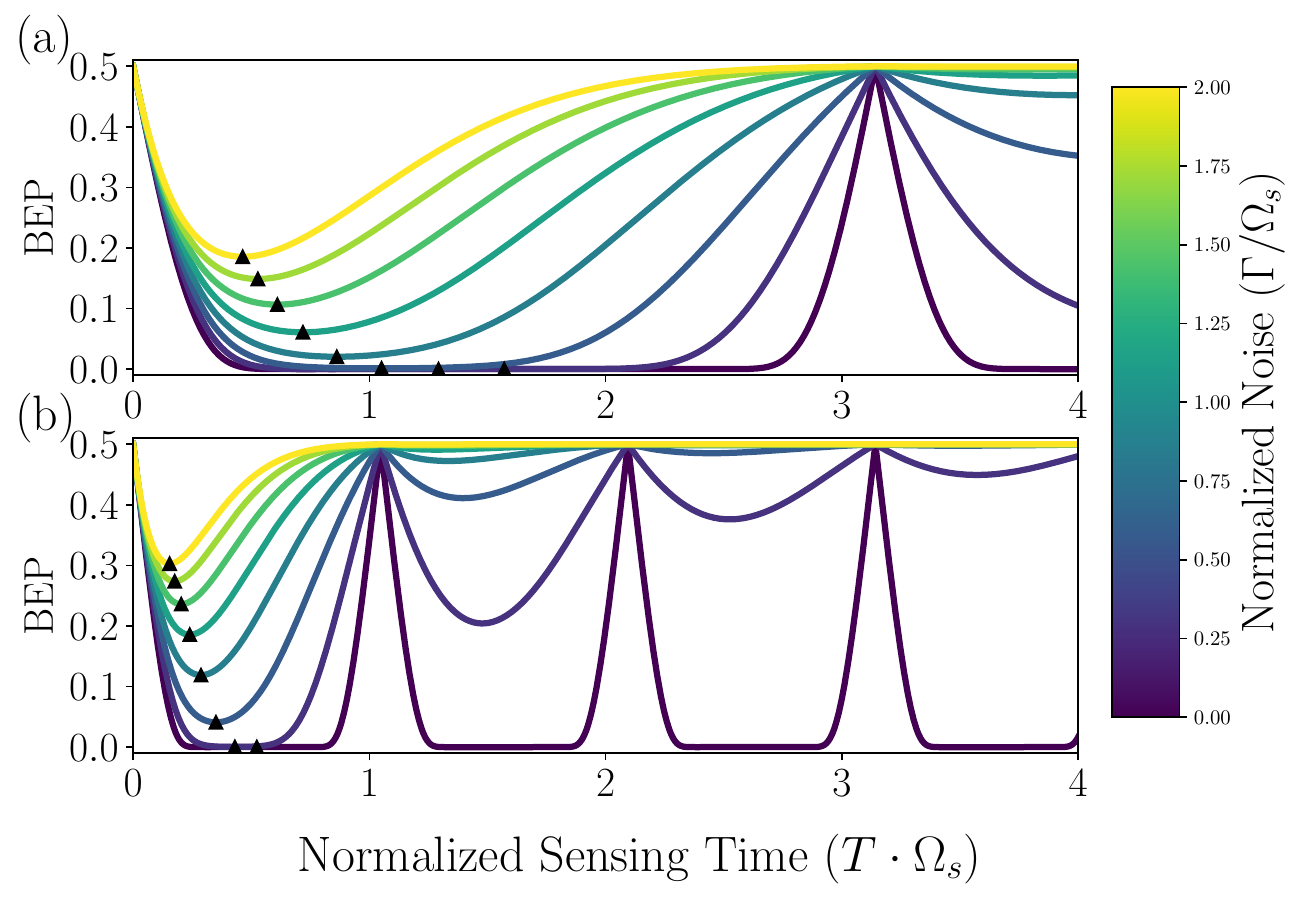}
\caption{
The optimal bit error probability (BEP) occurs for a sensing time inversely proportional to the signal strength, and with a white-noise model, this would be
$T_{opt} = \frac{1}{m\Omega_{\rm s}}\tan^{-1}(\frac{\Omega_{\rm s}}{\Gamma})$, where $m$ is the number of entangled qubits, $\Omega_{\rm s}$ is the signal strength, and $\Gamma$ is the dephasing rate. 
Panel (a) illustrates the BEP (\equationautorefname{\ref{eqn:bep_ensemble}}) as a function of sensing time for three unentangled quantum spins, for multiple levels of $\Gamma$, from $\Gamma=0$ to $2\Omega_{\rm s}$.
Panel (b) illustrates the BEP for three quantum spin entangled into a GHZ state.
The bottom panel demonstrates that collective sensing reduces the optimal sensing time, as the phase accumulates $m$ times faster, for $m$ spins entangled, but note also, that the GHZ state is more sensitive to the noise environment as well.
For slower bit rates, the unentangled ensemble would be superior, but for fast bit rates which limit the sensing time, an ensemble of entangled sensors prove superior. 
}
\label{fig:bit_error_dephasing}
\end{figure}

Next, we will discuss the more general scenario of using multi-qubit sensors for demodulating BPSK signals. First, we consider how an entangled multi-qubit system can improve discrimination of the encoded symbols in the noiseless case.
For a $m$ qubit sensor, we initialize in the GHZ state,
$\ket{\psi^{+}_{\rm GHZ}}=\frac{1}{\sqrt{2}}\big( \ket{+}^{\otimes m} + \ket{-}^{\otimes m} \big)$.
Then the sensor evolves in the present of the BPSK signal, using the Hamiltonian given by Eq.~\eqref{eq:transverse-H}. 
We get that
\begin{align}
    &e^{-i\tilde H_{\rm T}[b]T}\ket{\psi^{+}_{\rm GHZ}} \nonumber\\
    &=\cos\left(\frac{m\Omega_{\rm s}T}{2}\right)\ket{\psi^{+}_{\rm GHZ}}
    + (-1)^b i\sin\left(\frac{m\Omega_{\rm s}T}{2}\right)\ket{\psi^{-}_{\rm GHZ}}.
\end{align}
Now, we can obtain the difference operator,
\begin{align}
    &\rho_1(T) -\rho_0(T)  \nonumber \\
    &= -\sin\big(m\Omega_{\rm s} T\big) \Big(-i \ket{\psi^{+}_{\rm GHZ}}\bra{\psi^{-}_{\rm GHZ}} +\textrm{h.c.}    \Big).
\end{align}
This lets us design a measurement basis that optimally discriminates between the two phases.
The eigenstates are $\{\ket{\psi^+_{\rm GHZ}}\pm i\ket{\psi^-_{\rm GHZ}}\}$.
%
Therefore, it is clear that the BEP is minimized by measuring
\begin{equation}
    \mathcal{O} = \Big( i \ket{\psi^{+}_{\rm GHZ}}\bra{\psi^{-}_{\rm GHZ}} - i\ket{\psi^{-}_{\rm GHZ}}\bra{\psi^{+}_{\rm GHZ}} \Big).
\end{equation}
The enhancement from entangling $m$ qubits is evident by examining the BEP,
\begin{equation}
    P_{\rm e} = \frac{1}{2} - \frac{1}{2}\sin(m\Omega_{\rm s} T).
\end{equation}
The optimal time will be $T_{\rm{opt}}=\frac{\pi}{2m\Omega_{\rm s}}$, so the sensing takes $O(m^{-1}\tilde\Omega_{\rm s}^{-1})$ time instead of $O(\tilde\Omega_{\rm s}^{-1})$.
Collective Z-dephasing noise and phase noise on the sensors increases the error rate to
\begin{equation}
    \label{eqn:bep_ghz}
    P_{\rm e} = \frac{1}{2} - \frac{1}{2}e^{-2\chi(T)} \sin\left( m\tilde \Omega_{\rm s} T \right),
\end{equation}
where $\chi(T)$ corresponds to the decoherence of a GHZ state. 
The derivation follows from Sec.~\ref{sssec:single_qb_meas}, since the multi-qubit system maps onto a two-level subspace spanned by $\{\ket{\psi^+_{\rm GHZ}},\ket{\psi^-_{\rm GHZ}}\}$.
In the case of white noise, the $m$ qubit GHZ state will decay as $\chi(T)=m\Gamma T/2$. 
In Fig.~\ref{fig:bit_error_dephasing} (b) we examine the BEP for entangled states. 
We note that non-collective Z-dephasing noise is more complex in the multi-qubit case because it breaks the symmetry of the two-level GHZ system. We will leave the analysis of such complex noise sources for future work.

\subsection{Ensemble of Quantum Receivers}

Finally, we discuss using ensembles of quantum receivers to reduce BEP. 
In a multi-qubit platform, 
we consider a total of $n$ qubits, with each receiver consisting of $m$ entangled probes, meaning we have an ensemble of $k=n/m$ receivers.
Note that $m=1$ will then correspond to the scenario of $n$ unentangled sensors, that is, $n$ single-qubit sensors initialized in the product state~[Eq.~\eqref{eq:rho-prod}]. We make the simplifying (and practical) assumption that all sensors (single- or multi-qubit) are measured with the optimal measurement bases derived above in Sec.~\ref{subsec:opt-meas-basis}.
We assume that the optimal POVMs (for single qubit or single GHZ state) are measured simultaneously on each of the $k$ separable states, resulting in a classical dataset consisting of a total of $k$ outcomes, with $k_0$ `0' outcomes and $k_1$ `1' outcomes, $k_0+k_1=k$. 
The decision boundary for determining whether the encoded symbol is `0' is determined by the choice of an appropriate threshold $k^*$ of the number of `0' outcomes . The average BEP for this threshold corresponds to,
\begin{equation}
    P_{\rm e} = P( k_0 < k^* | 0) + P(k_0 > k^* | 1).
\end{equation} 
The minimum average BEP is achieved when $k^*$ is chosen such that the false positives and false negatives are minimized. For an unbiased communication channel (where `0' and `1' are equally likely to be transmitted) and with no preference towards false positives nor false negatives, the optimal choice is $k^*=k/2$.
For large ensembles, a binomial probability distribution can be approximated by a normal distributions under the central limit theorem. So in the large $k$ limit the BEP becomes,
\begin{equation}
    \begin{split}
    P_{\rm e} = \frac{1}{2}\int^{k^*}_{-\infty} P_\mathcal{N}(n; p_0 k, \sqrt{k p_0 p_1}) dn \\
    + \frac{1}{2}\int^\infty_{k^*} P_\mathcal{N}(n; p_1 k, \sqrt{k p_0 p_1}) dn.
    \end{split}
\end{equation}
We substitute in the definition for the individual bit errors to simplify and integrate. 
Therefore, we get the BEP in terms of the effective Rabi frequency, $\tilde\Omega_s$, sensing time, $T$, the decoherence parameter, $\chi(T)$, the number of qubits, $n$, and qubits per sensor, $m$,
\begin{equation}
    \label{eqn:bep_ensemble}
    P_e = \frac{1}{2}\mathrm{erfc}\Bigg(\frac{2e^{-2\chi(T)}\sin(m\tilde\Omega_s T)}{\sqrt{1-e^{-4\chi(T)}\sin^2(m\tilde\Omega_s T)}}\sqrt{\frac{n}{m}} \Bigg).
\end{equation}
Since \texttt{erfc} is a monotonically decreasing function, we can clearly see that the BEP decreases with more sensors $n/m$ in the ensemble and more phase accumulation $m\tilde \Omega_{\rm{s}}T$, but increases with stronger decoherence $\chi(T)$.

\section{Channel Capacity}\label{sec:channel_capacity}
Now, we derive the channel capacity for the various single- and multi-qubit quantum receivers. The channel capacity sets fundamental limits on the performance and data rate of the communication channel. We start by expressing the channel capacity in terms of the Holevo information. The Holevo bound establishes an upper bound on the amount of information that can be extracted from a quantum state~\cite{Holevo1973, Nielsen_Chuang_2010}. It bounds the maximum amount of information one can gain when a quantum channel is used to encode classical information. 
Suppose that the sender can prepare a fixed state $\{\rho_1, \cdots, \rho_l\}$, where the $j$th state is prepared with probability $p_j$.
The Holevo information bounds the classical mutual information, that is, the accessible information between the sent and received classical registers, maximized over the choice of measurement,
\begin{align}
    \mathrm{HI} = S\left( \sum_{j}^l p_j \rho_j \right) - \sum^l_j p_j S(\rho_j) \leq n
\end{align}
where $S(\rho)=-\Tr[\rho\log_2\rho]$ is the von Neumann entropy of the quantum state $\rho$ (in bits).
For a $l$-PSK communication channel, we will assume all symbols are equally likely, so that $p_j = 1/l$.
The channel is interrogated at a frequency, $1/T_{\mathrm{sym}}$, so we define the \emph{quantum channel capacity} as
\begin{align}
    C_{\mathrm{quantum}} = \max_{T_{\mathrm{sym}}} \frac{\mathrm{HI}}{T_{\mathrm{sym}}}.
\end{align}
In the noiseless case, the transmitted state $\rho_j$ is pure, so $S(\rho_j)=0$, otherwise, it will be a mixed state with non-zero von Neumann entropy.
The Holevo information is monotone under quantum channels, which means that the Holevo information of a noisy quantum channel is upper bounded by the noiseless Holevo information---this also follows from the data processing inequality.
Therefore, in this section, we will focus on calculating the noiseless quantum channel capacity as it bounds the noisy quantum channel capacity.
We compare the quantum channel capacity to the fundamental limit on classical channel capacity given by Eq.~\eqref{eqn:classical_capacity_limit}, and these comparisons are plotted in Fig.~\ref{fig:chu_limit}.

\subsection{Holevo Information for Product States\label{sec:single_qubit_sensing}}

Now, we will discuss the channel capacity of a quantum receiver made up of two-level quantum systems. While a quantum advantage is not anticipated with a single qubit receiver, the analysis developed here sets up our extension to ensemble sensors.

From Eq.~\eqref{eqn:transversal}, the single-qubit Hamiltonian corresponding to symbol $j$ of a $l$-PSK modulated signal is
\begin{align}
    H_j = \frac{\Omega_{\rm s}}{2}\left( \cos\phi_j \sigma^x + \sin\phi_j\sigma^y \right),
\end{align}
where the symbol is encoded in the phase $\phi_j = 2\pi j/l $ for $j=0, \cdots, l-1$. The corresponding unitary, after using Euler's formula, is
\begin{align}
    U_j(T) = \cos(\Omega_s T/2)\mathbb{I}
    - i\sin(\Omega_s T/2)(\cos\phi_j \sigma^x + \sin\phi_j \sigma^y).
\end{align}
It follows that the $j$th encoded single-qubit state is 
\begin{align}
    \ket{\phi_j}
    &=  \cos(\Omega_s T/2)\ket{0} - i\sin(\Omega_s T/2)e^{-i\phi_j}\ket{1}.
\end{align}
The density operator, averaged over the symbols, is
\begin{align}
    \bar\rho = \frac{1}{l}\sum^{l-1}_{j=0} \ket{\Phi_j}\bra{\Phi_j},
\end{align}
where we've defined the $n$-qubit state $\ket{\Phi_j} = \ket{\phi_j}^{\otimes n}$. Since $\ket{\Phi_j}\bra{\Phi_j}$ is a pure state, 
the ensemble spans a subspace of the Hilbert space $\mathcal{H}$ with dimension $\leq l$.
Therefore, there are at most $l$ nonzero eigenvalues. 
Consider a linear map $A: \mathbbm{C}^l \to \mathcal{H}$, where $A\ket{e_j}=\ket{\Phi_j}/\sqrt{l}$ and $\ket{e_j}$ is the standard basis on $\mathbbm{C}^l$.
It is straightforward to see that $\bar\rho = A A^\dagger$ and $M = A^\dagger A$, where $M$ is the $l \times l$ Gram matrix~\cite{Horn_Johnson_2012},
\begin{align}
    M_{jk} = \frac{1}{l}\braket{\Phi_j | \Phi_k} = \frac{1}{l}\braket{\phi_j | \phi_k}^n.
\end{align}
Then, the non-zero eigenvalues of $\bar \rho$ are exactly the eigenvalues of the Gram matrix.
Clearly the von Neumann entropy is maximized when the eigenvalues are all equal to $1/l$, so given a $l$-PSK scheme, $\mathrm{HI} \leq \log_2 l$.
And the BPSK case can be worked out exactly, since $\braket{\Phi_0 | \Phi_1} = \cos^n(\Omega_{\rm{s}} T)$, so
\begin{align}
    \mathrm{HI} = -\sum_{\alpha=\{\pm 1\}} \left( \frac{1+\alpha\cos^n(\Omega_{\mathrm{s}}T)}{2} \right)\log_2\left( \frac{1+\alpha\cos^n(\Omega_{\mathrm{s}}T)}{2} \right).
\end{align}
Unfortunately finding the Holevo information analytically for a large number of symbols, $l$, becomes impractical because it requires solving an $l\times l$ eigenvalue problem.

Instead, we use a different technique to calculate the Holevo information for an $n$ qubit sensor.
First, we define the $n$-qubit computational basis elements $\ket{x}=\ket{x_1 x_2 \cdots x_n}$ and $\ket{y}=\ket{y_1 y_2 \cdots y_n}$, where $x_i,y_i\in\{0,1\}$.
Then, the matrix element of $\ket{\Phi_j}\bra{\Phi_j}$ is 
\begin{align}
    & M_{x,y} = \braket{x|\Phi_j}\braket{\Phi_j|y} \nonumber\\
    &= \prod^n_{k=1} (\sqrt{p_0})^{1-x_k}(-i\sqrt{p_1}e^{-i\phi_j})^{x_k} (\sqrt{p_0})^{1-y_k}(i\sqrt{p_1}e^{i\phi_j})^{y_k} \nonumber\\
    &= (-i)^{|x|-|y|}\sqrt{p_0}^{2n-(|x|+|y|)}\sqrt{p_1}^{|x|+|y|} e^{-i\phi_j(|x|-|y|)}
\end{align}
where $p_0=\cos^2(\Omega_{\mathrm{s}}T/2)$, $p_1=\sin^2(\Omega_{\mathrm{s}}T/2)$, 
and $|x|$ is the Hamming weight of $x$. 
When we take the sum over all phases, $\phi_j$, we get
\begin{align}
    \sum^{l-1}_{j=0}e^{-i\phi_j(|x|-|y|)} = \sum^{l-1}_{j=0}e^{-i \frac{2\pi j}{l}(|x|-|y|)},
\end{align}
which is a finite geometric series
\begin{align}
    \sum^{l-1}_{j=0} r^j = \frac{1 - r^l}{1 - r} = \frac{1 - e^{-i2\pi(|x|-|y|)}}{1 - e^{-i \frac{2\pi(|x|-|y|)}{l}}}
\end{align}
where $r = e^{-i \frac{2\pi(|x|-|y|)}{l}}$.
This is zero unless $|x|-|y|$ is a multiple of $l$.
Therefore, for $l > n$, the off-diagonals of the matrix $M_{x,y}$ always vanish, and Eq. \eqref{eqn:multiqubit_entangled_channel_capacity} is diagonal in the computational basis.
If $l \leq n$, then some blocks survive, and there is coherence between states with Hamming weights differing by $l$.
The diagonal elements are
\begin{align}
    \braket{x | \bar\rho | x } = \frac{1}{l}\sum^{l-1}_{j=1} p_0^{n-|x|}p_1^{|x|} = p_0^{n-|x|}p_1^{|x|},
\end{align}
which is independent of $l$ and $j$ because the phases cancel out.
These are exactly the eigenvalues for $l>n$, with multiplicity $\binom{n}{|x|}$.
Therefore, the Holevo information is
\begin{align}
    \mathrm{HI}(l > n) = -\sum^n_{w=0} {\binom{n}{w}} p_0^{n-w}p_1^{w} \log_2\left( p_0^{n-w}p_1^{w} \right).
\end{align}
The Holevo information is independent of the PSK demodulation scheme as long as the number of symbols $l$ is greater than the number of qubits $n$.
Expanding the logarithm and simplifying, one finds that the sum depends only on the expected Hamming weight of the quantum strings. Explicitly,
\begin{align}
    \mathrm{HI}(l > n) = -\left[ \left(n-\mathbb{E}[W] \right)\log_2 p_0 + \mathbb{E}[W]\log_2 p_1 \right],
\end{align}
where $\mathbb{E}[W]$ is the average Hamming weight.
Under a binomial distribution, this reduces to
\begin{align}
    \label{eqn:holevo_info_unentangled}
    \mathrm{HI}(l > n) = n\left[-p_0\log_2 p_0 - p_1\log_2 p_1 \right] = n H_2(p_0),
\end{align}
showing that the Holevo information scales linearly with the number of qubits and the binomial entropy of the single-qubit distribution.
Because the Shannon entropy of the diagonals $S(\bar\rho_{\mathrm{diag}})$ is always greater than the von Neumann entropy $S(\rho)$, we can bound the Holevo information for $l \leq n$ by the Holevo information for $l > n$.
Therefore, we can bound the Holevo information for $l\leq n$ by the Holevo information for $l > n$.
Therefore, we can bound the quantum channel capacity
\begin{align}
    C_{\mathrm{quantum}}(T) \leq \max_{T_{\mathrm{sym}}} \frac{\min\left( n H_2\big(p_0(T)\big), \log_2 l\right)}{T_{\mathrm{sym}}},
\end{align}
where the sensing time is
\begin{align}
    T = \min \left(\frac{\pi}{2\Omega_{\mathrm{s}}}, T_{\mathrm{sym}}-T_{\mathrm{SPAM}}\right).
\end{align}
The channel capacity is modified to include phase noise and jitter by the substitution of the Rabi frequency $\Omega_{\mathrm{s}}\to\tilde\Omega_{\mathrm{s}}$, which reduces the quantum channel capacity.

\subsection{Multi-qubit Channel Capacity\label{sec:multiqubt_sensing}}

Next, we consider an ensemble of $n$ two-level quantum sensors, where $m$ qubits are entangled together per quantum receiver.
Naturally this assumes that $n$ is divisible by $m$.
As described previously in Sec.~\ref{ssec:initial_states}, the entangled sensor is initialized in the GHZ state $\ket{\psi^{+}_{\rm GHZ}}$.
As seen from Eq.~\eqref{eqn:bep_ghz}, entanglement increases the sensitivity of the sensor to the signal $m$-fold.
So, following the above derivation of the channel capacity, but noting the $m$-fold enhancement of Eq.~\eqref{eqn:bep_ghz}, we easily find the Holevo information for an ensemble of entangled sensors,
\begin{align}
    \mathrm{HI}(l >k) = k H_2(p_0)
\end{align}
where $k=n/m$, the number of independent measurements and $p_0 = \cos^2(m\tilde\Omega_\mathrm{s} T)$.
Therefore, the channel capacity of a quantum receiver is
\begin{align}
    \label{eqn:multiqubit_entangled_channel_capacity}
    C_{\mathrm{quantum}} \leq \max_{T_{\mathrm{sym}}} \frac{\min\left( kH_2(p_0), \log_2 l\right)}{T_{\mathrm{sym}}},
\end{align}
for demodulating an $l$-PSK signal with $k=n/m$ quantum receivers, each made up of $m$ entangled qubits.

By taking a Taylor expansion for $p_1 \ll 1/2$, about $p_1=0$, i.e, for a weak signal [$m\tilde\Omega_{\mathrm{s}}T\ll 1$], then
\begin{align}
    \mathrm{HI}(l>k)=-p\log_2 p + \frac{p}{\ln 2} + \mathcal{O}(p^2\log_2 p).
\end{align}
We find that
\begin{align}
    \mathrm{HI}(l>k) \approx \frac{n}{m}\left( -2(m\tilde\Omega_{\mathrm{s}}T)^2\log_2(m\tilde\Omega_{\mathrm{s}}T) + \frac{(m\tilde\Omega_{\mathrm{s}}T)^2}{\ln 2} \right)
\end{align}
So we can see that we maintain the  linear enhancement of Eq.~\eqref{eqn:holevo_info_unentangled} from having $n/m$ channels, but in addition, we get a quadratic enhancement due to the entanglement of $m$ qubits, which is analogical to theHeisenberg scaling in parameter estimation.

\section{Performance Comparison between Classical and Quantum Receiver\label{sec:comparison}}

\begin{figure*}[htb]
  \centering
  \includegraphics[width=\textwidth]{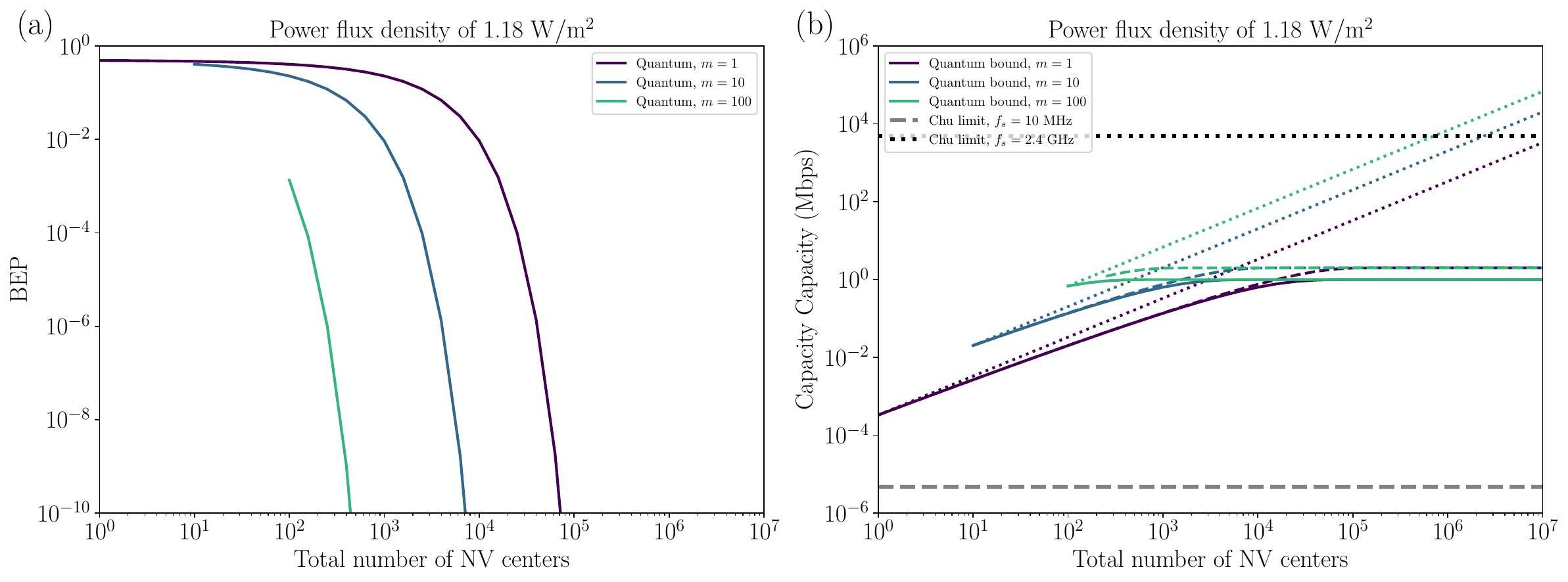}
  \caption{ 
Our quantum-enhanced sensing protocol can outperform the limits on classical electromagnetic receivers, because classical receivers become ineffective at small scales.
The receiver scale is set to 1 cm and the power flux density to 1.18 $\mathrm{W}/\mathrm{m}^2$, i.e., 100 nT at the quantum receiver.
In panel (a), BEP (\equationautorefname{\ref{eqn:bep_ensemble}}) is plotted against the total number of NV centers used in a quantum sensor.
Different traces are given for ensembles of quantum sensors initialized into GHZ states of $m$ spins, $\frac{1}{\sqrt{2}}\big( \ket{+}^{\otimes m}+\ket{-}^{\otimes m}\big)$.
We see that entangling the spins reduces the number of spins necessary for a given BEP by an order of magnitude.
In panel (b), the channel capacity for classical (Eq. ~\ref{eqn:classical_capacity_limit}) and quantum (Eq.~\ref{eqn:multiqubit_entangled_channel_capacity}) receivers are plotted against the total number of NV centers used in a quantum sensor.
The noiseless quantum channel capacities plotted are the $l>n$ bound (dotted), the QPSK encoding (dashed), and the BPSK encoding (solid). 
The noiseless quantum channel capacity sets an upper bound on the noisy quantum channel capacity.
We find that the crossover between the quantum channel capacity and the Chu-limited channel capacity at 2.4 GHz occurs between one and ten million NV centers at a power flux density of $\sim 1\,  \mathrm{W/m^2}$ at the receiver, depending on the amount of entanglement per quantum sensor.
The upper bound on the quantum channel capacityoutperforms the Chu limit on classical channel capacity at 10 MHz.
 }
  \label{fig:chu_limit}
\end{figure*}

For the herein-developed quantum receiver protocol, we obtained analytical expressions for BEP in Sec.~\ref{subsec:opt-meas-basis} and for channel capacity in Sec.~\ref{sec:channel_capacity}.
Now, we will assess the different quantum receivers using BEP, and we will also compare the performance of the quantum receivers to an electrically-small classical receiver of similar size using the channel capacity.
Since the linear dimensions of the active region~[$a=\left(3V/4\pi\right)^{1/3}$] of a quantum sensor is typically much smaller than the signal wavelength~[$\lambda_{\rm s}=c/f_s$], a classical antenna of similar volume is confidently in the electrically-small regime~[$2\pi a/\lambda_s\ll 1$].
Comparisons to an electrically-small classical receiver are a common benchmark in the Rydberg sensor literature~\cite{backes_2024, Elgee_2025, Cox2018}.
We make these comparisons keeping fixed the following parameters: (i) sensor volume ($V$), and (ii) signal characteristics (i.e., $f_{\rm s}$, $f_{\rm b}$ and BPSK modulation scheme).
In accordance with the IEEE 802.11 local area network (LAN) technical standard~\cite{ieee_wifi}, we set the data rate at 1 Mbps [$f_{\rm b}=10^6$ Hz] and the carrier frequency to 2.4 GHz [$f_{\rm{s}}=2.4\times 10^9$ Hz].

In Ref.~\cite{Cox2018}, the authors compare a Rydberg sensor to an electrically-small antenna with $50\ \Omega$ Johnson noise at room temperature and electric field of $0.4\ \mathrm{V/cm}$.
From that, we get a noise PSD $N_0/2 = 8.28\times 10^{-19}\ \mathrm{W/Hz}$.
We set the dimension $a=1$ cm for a realistic classical receiver, based on the active area used in similar comparisons~\cite{Cox2018, Barry_2024}.
We calculate the BEP and channel capacity over a range of signal intensities, from $10^{-4}\, \rm{W/m^2}$ to $100\, \rm{W/m^2}$ (i.e., magnetic field strengths in the range $1\,\rm{nT}$ to $1\, \mu\rm{T}$).
We use the effective aperture of a lossless isotropic, $A = \lambda^2/(4\pi)$, to calculate the signal power of the classical receiver, $P_{\rm{s}}=\rm{intensity}\times A$.

The comparisons in this study focus on using a NV defect in diamond as a quantum sensor, which has a magnetic coupling $\gamma_e = 2\pi\times 28$ GHz/T~\cite{Barry_2024}.
Then, a magnetic field of $\approx 1\ \mu T$ yields a Rabi frequency $\Omega_{\rm s} = 2\pi \times 30$ kHz, which is on the order of the field strength we can expect from magnetometry and sensing literature~\cite{zhengzhi_2023, Cox2018, Barry_2024, Wang2021, Magaletti_2022, Lamba_2024}.
We assume a spatially uncorrelated dephasing model [$S^{zz}_{ij}(\omega) = \delta_{ij} S_L(\omega)$ and $S^{\mu\nu}(\omega)=0$ for $\mu,\nu\neq z$],
where the spectrum of the spin-bath coupling can be assumed to be Lorentzian:
\begin{align}
    S_L(\omega) = \frac{\Delta^2 \tau_c}{\pi(1+(\omega\tau_c)^2)}
\end{align}
with coupling strength $\Delta = 30\ \mathrm{kHz}$ and correlation time $\tau_c = 10\ \mathrm{\mu s}$~\cite{Bar-Gill2012}.
We evaluate the decoherence parameter [Eq.~\eqref{eqn:z_dephasing_rate}] and get that
\begin{align}
    \chi(T) = \frac{\Delta^2\tau_c^2}{2\pi}\left(T/\tau_c + e^{-T/\tau_c} - 1 \right).
\end{align}
With a reasonable symbol time $T_{\mathrm{sym}} < 1\ \mu s$, we are in the short-time limit [$T \ll \tau_c$] and then $\chi(T) \approx \Delta^2 T^2/4\pi$.
We will use these parameters to give a sense on the performance limits of the quantum receivers developed  in this study and compare to the classical limit.

In Fig.~\ref{fig:chu_limit}, we compare classical and quantum receivers in the transverse sensing mode.
Even at the upper bound of our estimates ($10^7$ NV centers), the corresponding densities remain well within experimentally achievable ranges and below the thresholds at which fabrication, control, or dipolar interactions become limiting~\cite{Ishiwata_2017, barry_2020, zhou_2020_prx, doi:10.1126/sciadv.adg2080, orphal-kobin_2023, hughes_2025}.
The fundamental limit on classical channel capacity is given by Eq.~\eqref{eqn:classical_capacity_limit}.
The quantum channel capacity is given by Eq.~\eqref{eqn:multiqubit_entangled_channel_capacity} for an ensemble of $n/m$ independent sensors with $m$-qubit entangled (GHZ) states. 
The BEP for the ensemble of quantum sensors is given by Eq.~\eqref{eqn:bep_ensemble}.
We see that there is a quadratic enhancement of the quantum channel capacity with an $m$-qubit GHZ state.
Additionally, we demonstrate that by increasing the density of quantum spins per volume, one can exponentially reduces the bit error rate for a quantum-enhanced receive chain.
The classical channel capacity is bounded by the system volume, $C_{\rm{classical}}\leq \mathcal{O}\left(x^3\log(1/x)\right)$, where $x=ka$ and $x \ll 1$, i.e., the electrically-small regime.
While the quantum channel capacity is also proportional to volume [$C_{\rm{quantum}}\sim\mathcal{O}( \rho x^3)$], by increasing defect density, the quantum channel capacity bound increases linearly.
Furthermore, we can obviously see from Eq.~\eqref{eqn:classical_capacity_limit}, that the Chu-limited channel capacity is directly proportional to the signal frequency, $f_\mathrm{s}$. Of course, this limits the effectiveness of electrically-small antenna for low frequencies (i.e., long wavelengths), whereas there is no such restriction on the quantum protocol developed in this work.
Therefore, the quantum receiver protocol in the longitudinal mode, i.e., in the regime of low signal frequencies, has significant potential to outperform the classical limits on electromagnetic sensors.
On the other hand, the quantum channel capacity [Eq.~\eqref{eqn:multiqubit_entangled_channel_capacity}], is limited to symbol rates that are slower than the inverse SPAM time ($T_{\mathrm{sym}}>T_{\mathrm{SPAM}})$, which limits the maximal channel capacity.
One could hypothetically get around this limit by increasing the number of quantum sensors and cycling through them in a staggered fashion.
From our results, it is clear that a quantum BPSK receiver can surpass the fundamental limits of a classical receiver.
The caveat is that controlling and entangling enough qubits to beat the classical limits becomes a major engineering challenge.

\section{Error Suppression\label{sec:error_suppression}}

So far, we have laid out the demodulation protocol, an analysis of BEP with and without non-Markovian noise,  extended the protocol to multiple qubits, calculated channel capacity of quantum receivers, and made a comparison against the fundamental classical limit for PSK demodulation. 
Next, we turn towards error suppression. 
Our quantum sensing protocol relies on quantum control, so the full sundry of quantum control techniques can be applied to the system to suppress noise and errors: dynamical decoupling (DD) \cite{Viola_1999} and decoherence-free subspaces (DFS) \cite{lidar_2014}. In addition, error mitigation, such as zero-noise extrapolation (ZNE) \cite{Giurgica_Tiron_2020}, and quantum error correction codes (QECC) \cite{roffe_2019} could be implemented to increase signal fidelity. One of the strengths of this quantum-enhanced sensing protocol 
is the ability to draw upon interdisciplinary tools from digital signal processing, quantum sensing, and quantum computing. Since the above are all well-known and well-implemented techniques, we leave their implementation in our sensing protocol for future studies.

\begin{figure}[tb]
    \centering
    \includegraphics[width=\linewidth]{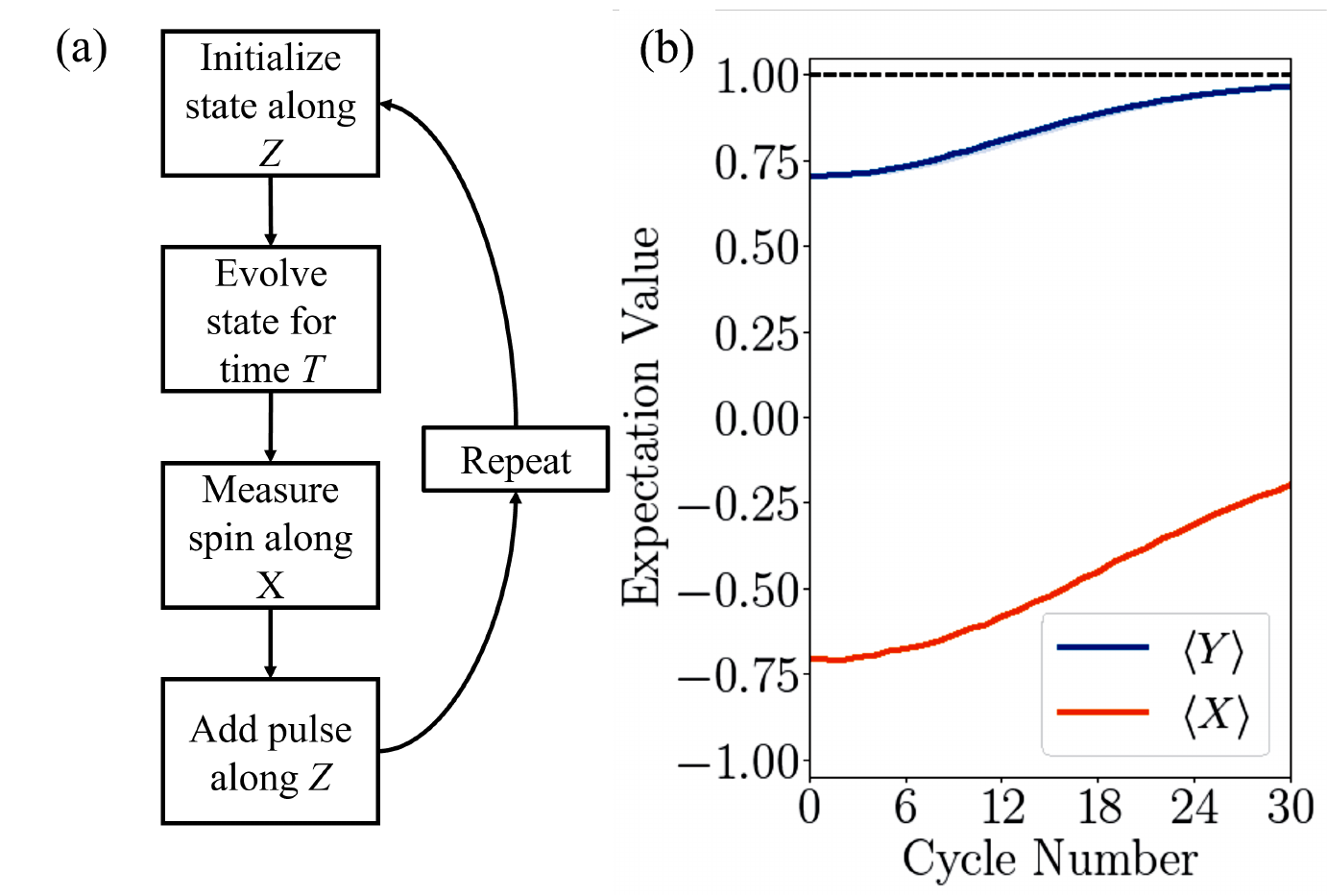}
    \caption{
A phase offset error occurs when the phase of an incoming signal is different from the phase of the `local oscillator' of the receiver. In a quantum context, this is a phase difference between a signal and the quantum control drive.
Panel (a) demonstrates a feedback protocol to correct for a phase offset error in our developed PSK quantum demodulation scheme.
The phase error is proportional to $\braket{\sigma^x}$, so by adding fast $Z$ pulses to the dynamics, with a strength proportional to $\braket{\sigma^x}$, one can get the control to be in phase with the incoming signal.
Panel (b) demonstrates that this protocol successfully reduces the phase offset, as seen by $\braket{\sigma^x}\to 0$ with the number of feedback cycles.
The demodulated signal is encoded in the $\braket{\sigma^y}$ measurement, so the error is reduced as that approaches the true value of $\pm 1$, i.e. encoding a `0' or `1' symbol, respectively.
In the plot, the expectation values are calculated with a Monte Carlo simulation of 50 noise trajectories. 
}
\label{fig:feedback}
\end{figure}

One important and unique source of error, whose suppression we will now discuss, is the phase offset error. The phase offset error is unique to phase-shift keying signals. Therefore, we develop a control feedback loop to minimize the mean phase offset in our quantum sensing protocol. 
The phase offset problem is linked to a synchronization error in classical PSK demodulation, and this analogy drives our innovation. 
As a quantum analogy, the following protocol essentially calibrates out an over-rotation about $Y$.
For transverse sensing, we find that a fast $Z$ pulse can synchronize the carrier wave to the control oscillator. The evolution generated by the spin-lock drive in the $k$th iteration of the feedback loop is
\begin{equation}
    U_0(0,T)=e^{-i(\omega_{\rm q} +\Omega_{\rm c})T\sigma^z/2}e^{-i\delta_k \sigma^z/2}.
\end{equation}
The feedback loop works as follows. A test qubit is evolved for time T and measured along the $x$-axis. The qubit is reset and allowed to evolve for another time $T$. But during this evolution, a fast $Z$ pulse is inserted proportional to the previously measured $\langle \sigma^x \rangle$. This is repeated until one finds that, $\langle \sigma^x \rangle \approx 0$, 
which implies that the mean phase offset $\mu \approx 0$. 
In this way, our technique is a kind of adaptive control, where
\begin{equation}
    \label{eqn:feedback}
    \delta_{k+1} = \delta_k -\eta\overline{\braket{\sigma^x}}
\end{equation}
is implemented to reduce $\braket{\sigma^x}$ in the next iteration.
The feedback loop protocol is illustrated in Fig.~\ref{fig:feedback}.
A first-order ordinary differential equation can be used to model this feedback loop and verify its correctness and stability, see App.~{\ref{app:error_suppression_modelling}}. There are only two fixed points, $\phi_{\rm{FP}} = \{0, \pi\}$, as $\phi(t)$ is periodic in intervals of $2\pi$. 
For longitudinal sensing, a fast $X$ pulse is inserted instead, and the measurement is along $z$-axis.
Our developed feedback loop protocol reduces the phase offset asymptotically towards zero using fast $Z$ pulses, synchronizing the transverse-coupled signal to the control drive.
The reduction of the phase offset $\mu$ improves the accuracy of demodulation [Eq.~\eqref{eq:q_bep_single_noise_offset}] and the channel capacity [Eq.~\eqref{eqn:multiqubit_entangled_channel_capacity}].

\section{Conclusion\label{sec:conclusion}}

\begin{table}
    \centering
    \begin{tabular}{ |c||c|c|c|c|  }
     \hline
     \multicolumn{5}{|c|}{Quantum Sensing Regimes} \\
     \hline
     Mode &  Carrier & Coupling & Control & Output \\
     \hline
     \hline
     Transverse & GHz & $\hat{X}$ &  $\hat{Z}$ & $\braket{\hat{Y}}$ \\
     \hline
     Longitudinal & $\leq$ MHz & $\hat{Z}$ & $\hat{X}$ & $\braket{\hat{Y}}$ \\
     \hline
    \end{tabular}
    \caption{Demarcation of the sensing regimes for the quantum demodulation protocol. Practical limits on the power of the quantum control drive dictates how far off the signal carrier frequency can be from either $0$ Hz or the spins' natural frequency. The detuning is directly proportional to control drive power. Depending on the sensing mode, the signal and control occur on complementary axes.}
    \label{tab:sensing_modes}
\end{table}
\begin{table}
    \centering
    \begin{tabular}{ |c||c|c|c|c|  }
     \hline
     \multicolumn{5}{|c|}{Sensor Ensemble with $n$ Qubits} \\
     \hline
     Mode &  Bitrate & per Sens. & No. Sens. & $T_{\rm{opt}}$ \\
     \hline
     \hline
     Unentangled & Slower & 1 &  $n$ & $\frac{1}{\Omega_s}\tan^{-1}\Big(\frac{\Omega_s}{\Gamma}\Big)$ \\
     \hline
     Entangled & Faster & $m$ & $n/m$ & $\frac{1}{m\Omega_s}\tan^{-1}\Big(\frac{\Omega_s}{\Gamma}\Big)$ \\
     \hline
    \end{tabular}
    \caption{
    A comparison of using single-qubit or multi-qubit entangled quantum sensors in PSK demodulation with a dephasing rate $\Gamma$. In the entangled case, each group of $m$ qubits are initialized 
    in the entangled GHZ state. This allows for faster data rates to be processed, since the optimal time sensing time goes as $T_{opt}\sim O(1/m)$, but at the expense of $1/m$ fewer total receivers. This reduction in total number of measurements can increase BEP.
    }
    \label{tab:ensemble_comp}
\end{table}

In this study, a quantum-enhanced receiver chain was developed to demodulate phase-shift keyed signals. 
First, we presented the model Hamiltonian for sensing and introduced our new decoding protocol.
Next, the projective operator-valued measures (POVMs) were found, which optimally discriminate between the state under the influence of a signal encoding the phase symbols---`1' and `0' in BPSK. 

We developed the protocol for two sensing modes: transverse and longitudinal coupling, and their benefits are summarized in Table~\ref{tab:sensing_modes}.
The benefits and problems with unentangled and entangled ensembles of quantum sensors was analyzed, an these finds are summarized in Table~\ref{tab:ensemble_comp}.
The most practical qubit platform to implement the protocol established in this study would be with NV defect centers in diamond. We believe that NVs in diamond would prove the best due to their optical properties and powerful defect engineering that allows for high density of controllable spins \cite{Ishiwata_2017, barry_2020, zhou_2020_prx, orphal-kobin_2023, hughes_2025}. 
Furthermore, the natural resonance of NV defect centers is in the range of 2 - 3 GHZ, which could be easily tuned to resonate with the 2.4 GHz carrier frequency of most PSK signals~\cite{ieee_wifi, wang2022}. Therefore, a nitrogen-vacancy defect platform would be the ideal candidate for a quantum-enhanced receiver chain to demodulate PSK signals in the limits beyond classical demodulation.

Next, we used a cumulant expansion to quantify the effects of temporally correlated dephasing noise and phase mismatch between the signal and control. 
The bit error probability (BEP) and an upper bound on channel capcacity was evaluated for the quantum protocol.
A comparison between this novel quantum receiver was made to an electrically-small classical receiver. 
By making use of the Chu limit, it was shown that the quantum receiver can achieve higher channel capacities than the limits of classical electromagnetic receivers.
We find that the quantum receiver can better outperform an electrically-small electromagnetic antenna in the low frequency, i.e., longitudinal, mode, compared to the high frequency, i.e., transversal, mode.
Finally, with an understanding of noise in the quantum sensing platform, a protocol was developed to suppress the phase offset error in a feedback loop, thereby reducing the BEP and increasing the channel capacity.
By using entangled probes for PSK demodulation, the developed protocol reaches the Heisenberg limit.

\section{Acknowledgements}
The authors would like to thank John van Dyke, Gregory Quiroz, Aidan Reilly, Kevin Schultz, Tim Sleasman, Ra’id Awadallah, and Nathaniel Watkins for useful discussions and feedback.
The authors also acknowledge funding from the Internal Research and Development program of
the Johns Hopkins Applied Physics Laboratory.

\appendix

\section{Quantum Demodulation of a General Phase-Shift Keying Signal\label{app:quantum_psk}}

The main study focuses on the demodulation of binary phase-shift keyed communication signals, but the method we develop is easily extensible. Here, it will be shown how to apply our methodology to receive quadrature phase-shift keyed signals (QPSK) as an example. 
For QPSK, the symbols are enumerated by $\phi(t) \in \{0, \pi/2, \pi, 3\pi/2\}$, where $\phi(t)$ encodes the signal into the carrier wave of Eq.~{\eqref{eqn:psk_signal}}.
For each phase, $\tilde H(t)$ is unique, therefore we should be able to be discriminated using the right measurements.

Like in the main text, we will work in transverse-sensing mode to illustrate the ease of general PSK demodulation with our quantum protocol.
First, consider that for $\phi \in \{0, \pi\}$, the toggling-frame Hamiltonian [Eq.~{\eqref{eqn:transversal}}] is only along the x-axis, and the resultant evolution rotates the Bloch vector about the x-axis. If the system is prepared in a $Z$ eigenvector, then it would be optimal to measure $\sigma^y$.
But if $\phi \in \{\pi/2, 3\pi/2 \}$, the rotation is about the y-axis.
In that case, measuring $\sigma^y$ is meaningless, but as the state would rotate in the $xz$ plane, one can easily show that a $\sigma^x$ measurement would be optimal.
Therefore by preparing the system in a $\sigma^z$ eigenvector and measuring along the $x$- and $y$-axes, we can completely discriminate the phase. 
Measurements along $y$ are in-phase components of the signal, and measurements along $x$ are quadrature components,
\begin{equation}
    \begin{split}
    \braket{\sigma^y(T)} &= \Tr \Big[ \frac{1}{2}\Big( I + e^{-i\int^T_0 \tilde H(t)dt}\big(\hat x \cdot \vec \sigma \big) e^{i\int^T_0 \tilde H(t)dt}\Big) \sigma^y \Big] \\
    &= \frac{1}{2} \sin(\Omega_s T) \cos\phi,
    \end{split}
\end{equation}
\begin{equation}
    \begin{split}
    \braket{\sigma^x(T)} &= \Tr \Big[ \frac{1}{2}\Big( I + e^{-i\int^T_0 \tilde H(t)dt}\big(\hat x \cdot \vec \sigma \big) e^{i\int^T_0 \tilde H(t)dt}\Big) \sigma^x \Big] \\
    &= \frac{1}{2}  \sin(\Omega_s T) \sin\phi,
    \end{split}
\end{equation}
where we have synchronized the measurements so that $U_0(T)=I$.
From these two measurements we can reconstruct an IQ diagram, like in classical demodulation of PSK signals.
Thus, by doubling the number of measurements, we can extend our methodology to any PSK modulation scheme. For instance, we could apply this to any variant of QPSK (offset QPSK, $\pi/4$-QPSK, dual-polarization QPSK), 8-PSK, 16-PSK, or quadrature amplitude modulation (QAM).

\begin{figure}
    \centering
    \includegraphics[width=0.9\linewidth]{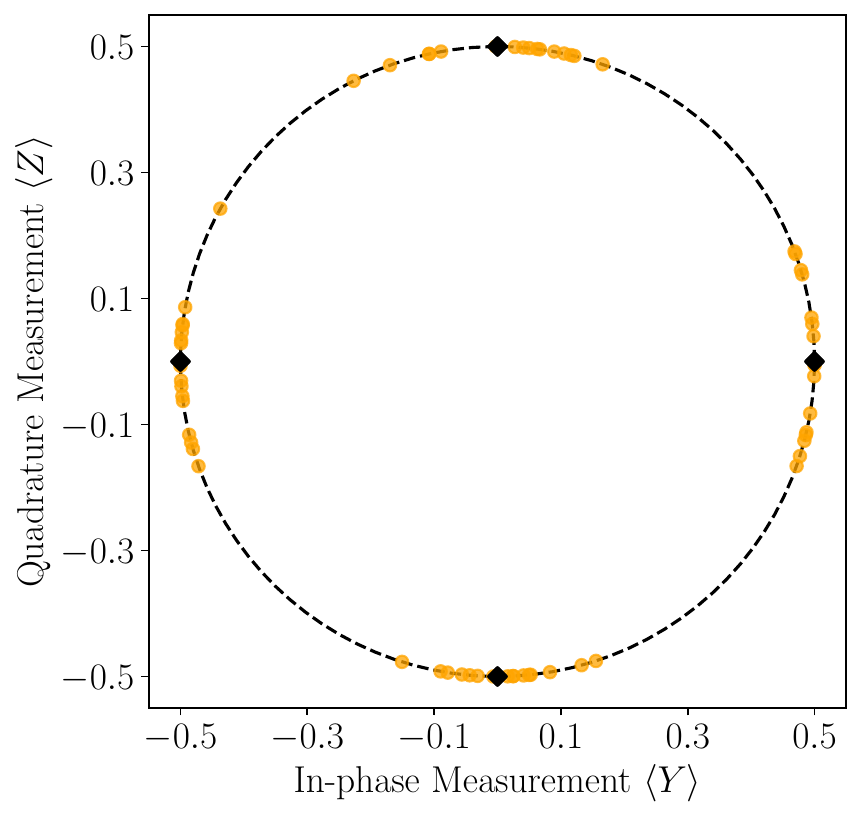}
    \caption{IQ Diagram demonstrating the demodulation of quadrature phase-shift keying (QPSK) signal with a quantum sensor in the transverse-sensing mode. Measurements along $\sigma^y$ and $\sigma^x$ mimic the in-phase and quadrature-phase of a traditional IQ diagram.}
    \label{fig:qpsk_quantum_example}
\end{figure}

\begin{figure}[t]
    \centering
    \includegraphics[width=0.8\linewidth]{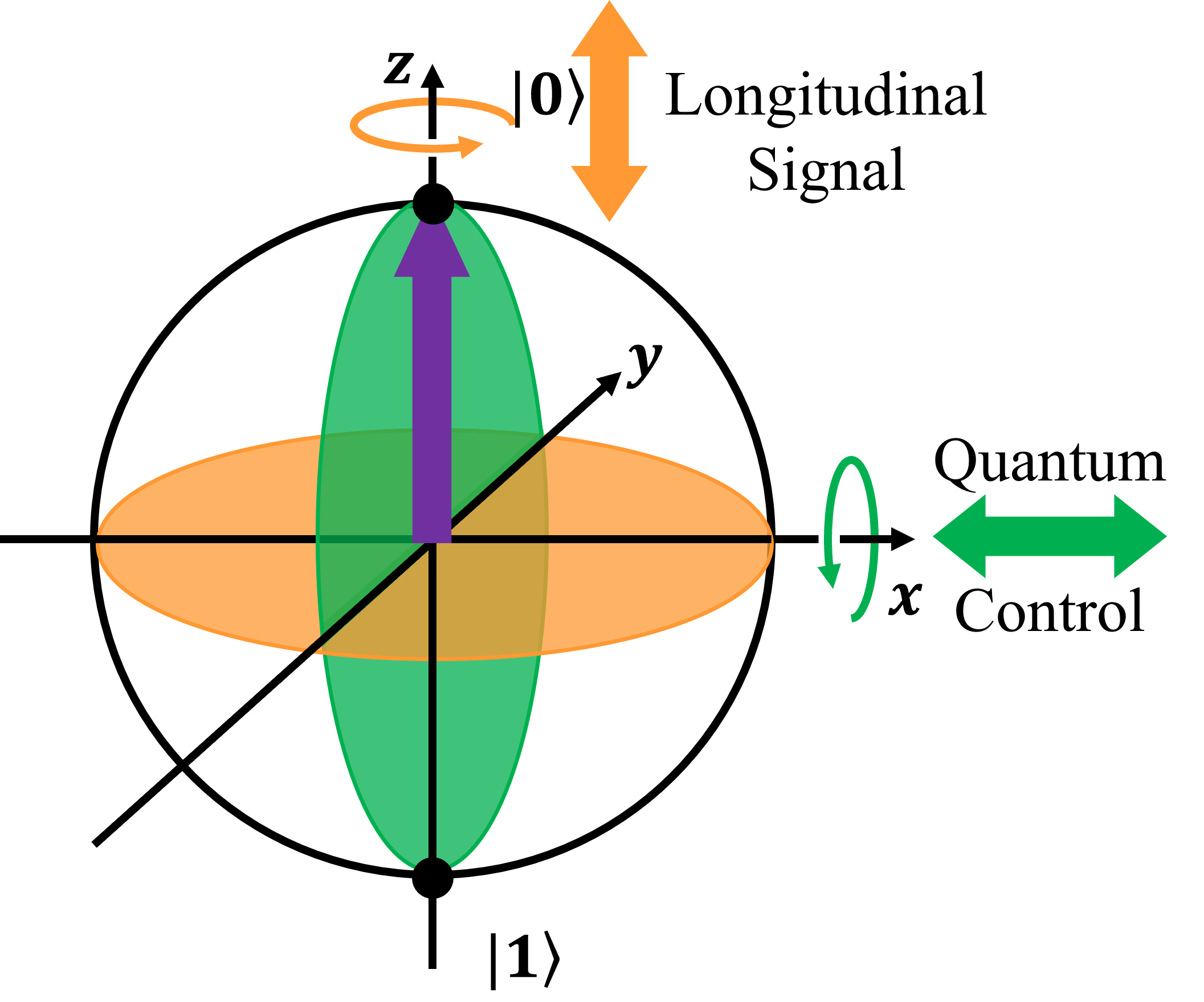}
    \caption{Schematic representation of a qubit in the quantum demodulation protocol.
    A sensor is prepared the ground state $\ket{+}^{\otimes n}$ and measured along the $y$-axis to demodulate the signal. The orange arrow denotes the direction of the signal coupling, the green arrow the control axis, and the purple arrow the initial state.
    Longitudinal coupling of the signal requires constant transverse control.
    }
    \label{fig:sensing_modes_longitudinal}
\end{figure}

\section{Longitudinal coupling}\label{subsubsec:control for longitudinal}

Longitudinal coupling between the signal and the receiver is advantageous when the signal frequency is close to the \emph{Rabi frequency} of the applied control. 
We work under the assumption that the frequencies satisfy, $\omega_{\rm q}\gg \omega_{\rm s}\approx\Omega_{\rm c}$.
Consider a transverse control drive (along $x$ axis, $\Omega^x=\Omega_c$, $\Omega^z=0$) resonant with the qubit frequency,
$\vec\Omega(t)=\left(\Omega_c\cos(\omega_q t),0,0\right)$.
The receiver Hamiltonian in the lab frame is given by
\begin{equation}
    \label{eqn:longitudinal_lab_frame}
    H_{\rm L}(t) = \sum^n_i \Big( \frac{\omega_{\rm q}}{2}\sigma^z_i +\Omega(t)\sigma^x_i+s(t)\sigma^z_i + \vec\eta_i(t)\cdot\vec\sigma_i \Big),
\end{equation}
where the Rabi frequency is given by $\Omega_{\rm c}$.
We apply Eq.~{\eqref{eqn:toggling_frame}} to longitudinal coupling.
So the relevant control matrix elements for the signal are
$R_i^{zz}(t)= \cos\Omega_c t$
and
$R^{zy}_i(t)= \sin\Omega_c t$.
After some multiplicative trigonometric identities, the toggling-frame Hamiltonian is
\begin{widetext}
\begin{equation}
    \begin{split}
    \tilde H_{\rm L}(t) = \sum^n_i
    \Bigg( \frac{\Omega_{\rm s}}{2} \Big( 
    \Big[ \cos\big((\omega_s+\Omega_c)t+\phi(t)\big) 
    +\cos\big((\omega_s-\Omega_c)t+\phi(t)\big) \Big]\sigma^z_i \\
    +\Big[ \sin\big((\omega_s+\Omega_c)t+\phi(t)\big) 
    +\sin\big((-\omega_s+\Omega_c)t-\phi(t)\big) \Big]\sigma^y_i 
    +  \sum_{\mu\nu\in\{x,y,z\}}\eta^\mu_i(t) R^{\mu\nu}_i(t)\sigma^\nu_i \Bigg).
    \end{split}  
\end{equation}
\end{widetext}
When the Rabi frequency of the drive matches the signal frequency, $\Omega_{\rm c}=\omega_{\rm s}$, also known as \emph{spin-locking}, 
and under the rotating-wave approximation (RWA), $\omega_s T_{\rm{sym}} \gg 1$,
this Hamiltonian 
becomes
\begin{equation}
    \label{eqn:longitudinal}
    \begin{split}
    \tilde H_{\rm L}(t) = \sum^n_i\Bigg( \frac{\Omega_{\rm s}}{2}\Big( \cos\phi(t) \sigma^z_i + \sin\phi(t) \sigma^y_i \Big) \\
    +  \sum_{\mu\nu\in\{x,y,z\}}\eta^\mu_i(t) R^{\mu\nu}_i(t)\sigma^\nu_i \Bigg),
    \end{split}
\end{equation}
where again the noise is transformed, as described in \sectionautorefname{\ref{ssec:frame_transforms}}. 
Likewise, it is clear that the effective evolution of the qubit state in the toggling-frame is determined by symbol encoded in the signal phase. 

\section{Derivation of optimal BEP for noisy single-qubit with transverse signal\label{app:noisy_optimal_measurement}}

Sec.~{\ref{subsec:opt-meas-basis}} covers the optimal measurement for a single-qubit sensor with Z-dephasing, i.e. additive Gaussian noise along the z-axis. The main text focuses on a transverse signal, but these calculations can easily be applied to the longitudinal mode.
The control matrix in the transverse mode is quite simple, as the two frame transforms have the same axis of rotation,
\begin{equation}
    R^{\mu\nu}_i(t) = \begin{pmatrix}
        \cos\big((\omega_q+\Omega_c)t\big) & \sin\big((\omega_q+\Omega_c)t\big) & 0 \\
        -\sin\big((\omega_q+\Omega_c)t\big) & \cos\big((\omega_q+\Omega_c)t\big) & 0 \\
        0 & 0 & 1
    \end{pmatrix}.
\end{equation}
The second cumulant moment and decoherence parameters are given by Eq.~{\eqref{eqn:general-2nd-cumulant-1}} and Eq.~{\eqref{eqn:general-2nd-cumulant-3}}, respectively.
Assuming that the noise model is pure Z-dephasing, $S^{\gamma\delta}(\omega)=0$ except for $\gamma=\delta=z$. Then from the control matrix, it is clear that $\chi^{\mu\nu}(T)=0$ and $\psi^{\mu\nu}(T)=0$ except for $\mu=\nu=z$.
The dephasing rates  for $\chi^{zz}(T)$ and $\psi^{zz}(T)$, are given by Eq.~{\eqref{eqn:general-2nd-cumulant-4}} and Eq.~{\eqref{eqn:general-2nd-cumulant-5}.
Now the second term falls out because the associated operator is zero, 
$[\mathcal{B}_\mathcal{O}]^{zz}=\mathcal{O}[\sigma^z,\sigma^z]\mathcal{O}^{-1}-[\sigma^z,\sigma^z]=0$.

Also since
\begin{align}
    [\mathcal{A}_\mathcal{O}]^{zz} &= 2I - 2\mathcal{O}^{-1}\sigma^z\mathcal{O}\sigma^z,
\end{align}
then $[\mathcal{A}_\mathcal{O}]^{zz}=4I$ or $0$, depending on the observable $\mathcal{O}$.
The second cumulant can be separated from the the first cumulant, amounting to a scalar exponential factor for the observable,
\begin{equation}
    \Lambda(T)=e^{\mathcal{C}(T)}=e^{-2\chi(T)} e^{\mp i \Omega_s T \big( \sigma^x - \mathcal{O}^{-1}\sigma^x\mathcal{O}\big)/2}.
\end{equation}
We find that $\mathcal{O}\neq \sigma^x$, otherwise the first cumulant (encoding the phase information) would be zero. Then, we have the first cumulant simplify, $\sigma^x - \mathcal{O}^{-1}\sigma^x\mathcal{O}=2\sigma^x$, so
\begin{equation}
    \Lambda(T)=e^{-2\chi(T)} e^{\mp i \Omega_s T \sigma^x}.
\end{equation}
Then using the fact that $\mathcal{O}$ and $\sigma^x$ anti-commute,
\begin{equation}
    \overline{\braket{\mathcal{O}}} = e^{-2\chi(T)} \Tr\Big[e^{\mp i \Omega_s T \sigma^x/2} \rho_0 e^{\pm i \Omega_s T \sigma^x/2}\mathcal{O}\Big]
\end{equation}
Thus we arrive at the same dynamics as in the noiseless case, with the addition of an exponential decay term. Just as before, it can be shown that $\mathcal{O}=\sigma^y$ is the optimal measurement. 
The expectation can be plugged back into the Helstrom bound to get the bit error rate with dephasing noise along the z-axis [Eq.~{\eqref{eq:q_bep_single_noise_offset}}].

\section{Analytic Derivation of Fixed Points for Phase Error Suppression, \texorpdfstring{Eq.~{\eqref{eqn:feedback}}}{Eq. 62}\label{app:error_suppression_modelling}}

We examine the dynamics of the $\mathcal{O}\equiv\sigma^x$ observable in the presence of phase noise to understand the phase error suppression developed in the main text.
We introduce a fast pulse, $\delta$, along the z-axis during sensing to suppress the phase noise offset. This pulse alters the frame transformation
\begin{equation}
    U_0(T) = \exp[-i( (\omega_q+\Omega_c)t + \delta)\sigma^z/2],
\end{equation}
such that under RWA, the rotating-frame Hamiltonian changes from Eq.~{\eqref{eqn:transversal}} to
\begin{equation}
    \begin{split}
      \tilde H_{\rm T}(t) = \sum^n_i\Bigg( \frac{\Omega_{\rm s}}{2}\Big( \cos(\phi(t)+\delta) \sigma^x_i + \sin(\phi(t)+\delta)\sigma^y_i \Big) \\
      +  \sum_{\mu\nu\in\{x,y,z\}}\eta^\mu_i(t) R^{\mu\nu}_i(t)\sigma^\nu_i \Bigg).
    \end{split}
\end{equation}
We truncate the cumulant expansion at the first non-zero order which, for the case of phase errors, is the first cumulant,
\begin{equation}
    \mathcal{C}^{(1)}_\mathcal{O}(T) = 
    -i \Omega_{\rm s} T e^{-\sigma^2/2}\sin(\mu+\delta)\sigma^y.
\end{equation}
The observable-dependent error operator is $\Lambda(T)=e^{\mathcal{C}^{(1)}_\mathcal{O}(T)}$, so the noise-averaged expectation is
\begin{equation}
    \overline{\braket{\sigma^x}} = \Tr[\rho_0 \sigma^x e^{-i\alpha\sin(\mu+\delta)\sigma^y}],
\end{equation}
where for brevity we designate $\alpha = \Omega_{\rm s}T e^{-\sigma^2/2}$. Since $\sigma^x$ and $\sigma^y$ anti-commute and using trace properties, we may rearrange the terms such that
\begin{equation}
    \overline{\braket{\sigma^x}} = \Tr[e^{-i\alpha\sin(\mu+\delta)\sigma^y/2} \rho_0 e^{i\alpha\sin(\mu+\delta)\sigma^y/2} \sigma^x ].
\end{equation}
And since the initial state is $\rho_0 = \frac{1}{2}(I+\sigma^z)$, we may simplify our expressions further,
\begin{equation}
    \overline{\braket{\sigma^x}} = \frac{1}{2}\Tr[\cos(\alpha\sin(\mu+\delta))\sigma^z\sigma^x +\sin(\alpha\sin(\mu+\delta))\sigma^x \sigma^x],
\end{equation}
and finally using Pauli relations, we arrive at the noise-averaged expected value of $\sigma^x$ in terms of the mean and variance of the phase offset,
\begin{equation}
    \label{eqn:x_measurement}
    \overline{\braket{\sigma^x}} = \sin\left(\Omega_s T e^{-\sigma^2/2}\sin\left(\mu+\delta\right)\right),
\end{equation}
which is analogous to Eq.~{\eqref{eqn:y_measurement}} in the main text.

With our feedback protocol, we update the amplitude of the fast pulse in the next iteration, proportionally to the expected value of $\sigma^x$,
\begin{equation}
    \delta_{k+1} = \delta_{k} - \eta \overline{\braket{\sigma^x}}.
\end{equation}
After substituting in Eq.~{\eqref{eqn:x_measurement}}, we see that this is a difference equation 
which can be modeled as a non-linear first-order ordinary differential equation (ODE),
\begin{equation}
    \label{eq:non_linear_ode}
    \frac{d\delta}{dk} = -\eta\sin\left(\alpha\sin\left(\mu+\delta\right)\right).
\end{equation}
Clearly, there are fixed points at $\mu+\delta=n\pi$ for all $n\in\mathbb{Z}$.
For $|\mu+\delta| \ll 1$, i.e. around the fixed point $n=0$, Eq.~{\eqref{eq:non_linear_ode}} becomes approximately a linear first-order ODE,
\begin{equation}
    \label{eq:linear_ode}
    \frac{d\delta}{dk} \approx -\eta\alpha(\mu+\delta),
\end{equation}
and as such, the phase offset moves asymptotically to the fixed point,
\begin{equation}
    |\mu+\delta| \propto e^{-\eta\alpha k}
\end{equation}
as a function of the number of iterations $k$.
Therefore, with this feedback protocol, the phase offset decays exponentially to zero.

\end{document}